\documentstyle[eqsecnum,aps,twocolumn,prd]{revtex}
\input epsf
\def\plotone#1{\centering \leavevmode
\epsfxsize= 0.8\columnwidth \epsfbox{#1}}

\def\plotonecita#1{}
\def\plotonecfpa#1{\plotone{#1}}
\def\epsfboxcita#1{}
\def\epsfboxcfpa#1{\epsfbox{#1}}

\def\be{\begin{equation}}
\def\ee{\end{equation}}
\def\bea{\begin{eqnarray}}
\def\eea{\end{eqnarray}}

\def\cmm2{{\,\rm cm^{-2}}}
\def\cm2{{\,{\rm cm}^2}}
\def\cmm3{{\,{\rm cm}^{-3}}}
\def\gcmm3{{\,{\rm g\,cm^{-3}}}}

\def\VEV#1{\left\langle #1\right\rangle}
\def\fun#1#2{\lower3.6pt\vbox{\baselineskip0pt\lineskip.9pt
  \ialign{$\mathsurround=0pt#1\hfil##\hfil$\crcr#2\crcr\sim\crcr}}}
\def\C{{\cal C}}

\def\muK{\mu {\rm K}}

\def\'{^{\prime}}
\def\avrg#1{{\langle #1 \rangle}}

\def\eps{\varepsilon}
\def\pomega{\varpi}

\def\etal{{et al.}}

\def\half{{\textstyle{1\over2}}}

\def\p3m{P$^3$M}

\def\Tr{{\rm Tr}\,}

\begin{document}
\twocolumn[\hsize\textwidth\columnwidth\hsize\csname @twocolumnfalse\endcsname
\preprint{CfPA/97-th-11,CITA-97-31}
\title{Estimating the Power Spectrum of\\ 
the Cosmic Microwave Background}
\author{J.~R.\ Bond$^1$, A.~H.\ Jaffe$^2$, and L.\ Knox$^1$}
\address{$^1$ Canadian Institute for Theoretical Astrophysics, Toronto, ON M5S 3H8, CANADA}
\address{$^2$ Center for Particle Astrophysics,
  301 LeConte Hall, University of California, Berkeley, CA 94720}
\date{\today}
\maketitle

\begin{abstract}
  We develop two methods for estimating the power spectrum, $C_\ell$, of
  the cosmic microwave background (CMB) from data and apply them to the
  COBE/DMR and Saskatoon datasets.  One method involves a direct
  evaluation of the likelihood function, and the other is an estimator
  that is a minimum-variance weighted quadratic function of the data.
  Applied iteratively, the quadratic estimator is not distinct from
  likelihood analysis, but is rather a rapid means of finding the power
  spectrum that maximizes the likelihood function.  Our results bear
  this out: direct evaluation and quadratic estimation converge to the
  same $C_\ell$s.  The quadratic estimator can also be used to directly
  determine cosmological parameters and their uncertainties.  While the
  two methods both require $O(N^3)$ operations, the quadratic is much
  faster, and both are applicable to datasets with arbitrary chopping
  patterns and noise correlations.  We also discuss
  approximations that may reduce it to $O(N^2)$ thus making it practical
  for forthcoming megapixel datasets.
\vskip 0.2in \end{abstract}
]\pacs{Valid PACS appear here.}
\section{Introduction}

Observations of the cosmic microwave background (CMB) anisotropy
are providing strong constraints on theories of cosmological
structure formation.  Planned observations have the potential
of providing constraints on the parameters of these theories 
at the percent level\cite{knox95,forecast,bet97}.  

Predictions of theories for CMB anisotropy are statistical in nature.
For many theories, the complete description is given by
the power spectrum, $C_\ell$, defined below.  Thus extraction of $C_\ell$
from the data is of utmost importance as an end in itself and for
purposes of ``radical compression''\cite{ourotherpaper,jkb}.

With the assumption of the Gaussianity of the data, the likelihood
function---the probability of the data given a particular theory---takes
a simple form; with the further assumption of a prior uniform in the
parameters, the likelihood is proportional to the posterior distribution
of the parameters, given the data.  
This is precisely the quantity one wants and thus likelihood analysis
has been used extensively for calculating the constraints on
parameters given by CMB data.  This is true whether the parameters
are those of the power spectrum itself or cosmological parameters.

Another approach has been to form estimators that are quadratic
functions of the data, {\it e.g.}, \cite{HauPeebWright}.  
This procedure has been improved
recently by the use of minimum-variance weighting of all the pairs of
data points \cite{tegmarkoptimal,kbj}.  In this paper we present a
unification of the quadratic and likelihood approaches.  We show that,
when used iteratively, the minimum-variance weighted quadratic
estimator is a fast technique for finding the maximum of the
likelihood function.  

In Section II we introduce the likelihood function, explain our method
for evaluating it directly, and derive the quadratic estimator.  We
apply quadratic estimation and direct evaluation to the case of
COBE/DMR\cite{DMR} in Section III.  Both methods involve iteration and
we find that for both, the iteration converges rapidly, with excellent
agreement between the two methods on the final $C_\ell$s and their
variances.  However, the higher moments of the probability distribution
cannot be estimated with the quadratic approach---and we find that there
are significant deviations from Gaussianity in the likelihood as a
function of $C_\ell$.  We discuss these differences, problems arising
from them and possible solutions.

For COBE/DMR we estimate every individual $C_\ell$ (for $2 \le \ell \le 24$)
since the data allow us to determine these with some precision.
The quadrupole, $C_2$, has received more attention in
previous work than any of the other moments because of its small value
and because it is the most susceptible to contamination by
emission from our galaxy\cite{dmrFGs}.  We also find the quadrupole to
be quite small, $C_2 = 149 \pm 126 \ \muK^2$, compared to $C_2 = 810 \ 
\muK^2$ for COBE-normalized standard cold dark matter (CDM).  However, due to the strong
skewness of the probability distribution for $C_2$, 25\% of the
probability is actually above the COBE-normalized CDM value of $C_2$.
Thus consistency with relatively flat models like standard CDM does not
require the quadrupole power to have been reduced by systematic errors.

For most observations, which only cover a small fraction of the sky,
estimating every $C_\ell$ is not possible.  One must be content with
estimating the power spectrum either with some binning in $\ell$ or
through some other parameterization.  Therefore in Section IV we
discuss binning and rebinning.  Then in Section V we apply the methods
to estimate, from the Saskatoon (SK) data \cite{nett95}, 
the power in ten bins from $\ell = 19$ to $\ell = 499$.

Power spectrum estimation can be used as a form of data compression
where the estimates of $C_\ell$ and their covariance matrix are then
used to constrain cosmological parameters.  Because of the great
simplifications involved in working with power spectrum estimates
instead of pixelized data, this is currently the only practical
procedure for using all the CMB data.  Such exercises have been
conducted, {\it e.g.}, \cite{Lineweaver,BondJaffe,HancockRocha}. 
In Section
VI we discuss the approximations involved in such a procedure and
methods for reducing the resulting inaccuracies, and
in Section VII we apply these results to future balloon- and
satellite-borne experiments.

Unfortunately, direct evaluation of the likelihood function is an
$O(N^3)$ operation, where $N$ is the number of data points.  
And it must be evaluated many times.
Thus for $N \gtrsim 10,000$ this
procedure becomes rapidly intractable on modern workstations---at
least for the most straightforward implementations.
Although the speed of likelihood analysis has been greatly
increased by use of signal-to-noise eigenmode compression 
\cite{BondJaffe,Bond,Bond94,BunnWhite,TTH}, this procedure still
requires an $O(N^3)$ operation to be performed at least once.

Further speed is necessary if we are to be able to analyze forthcoming
megapixel datasets.  The quadratic estimator may offer a means
of achieving this speed.  We emphasize that as we have applied it
here it is still an $O(N^3)$ operation, but believe that approximations
may be made in a controlled manner to reduce it to $O(N^2)$.  
We discuss these problems and possible solutions in Section VIII, as well
as explicitly outline our algorithm for power spectrum estimation from
CMB data.

\section{Methods: Likelihood Analysis}
\label{sec:methodsLike}

We begin by establishing the notation used for
describing the pixelized data of a CMB observation.  
We also define the power spectrum, $C_\ell$, and the likelihood
function.  With this common groundwork complete, we then move on to
a description of the two different methods for estimating $C_\ell$.

\subsection{The Likelihood Function}
In general, CMB observations are reduced to a set of
binned observations of the sky, or pixels, $\Delta_i$,
$i=1\ldots N$
together with a noise covariance matrix, $C_{nii'}$.
We model the observations as contributions from signal and noise,
\be
\Delta_i = s_i+n_i.
\ee
We assume that the signal and noise are
independent with zero mean, with correlation matrices given by
\begin{equation}
  C_{Tii'}= \langle s_i s_{i'} \rangle;\quad
  C_{nii'}= \langle n_i n_{i'} \rangle 
\end{equation}
so
\begin{equation}
  \langle \Delta_i \Delta_{i'} \rangle = C_{Tii'}+C_{nii'}
\end{equation}
where $\langle\ldots\rangle$ indicate an ensemble average.
With the further assumption that the data are Gaussian, these
two point functions are all that is necessary for a complete
statistical description of the data.

One important complication to the above description comes from the
existence of constraints.  Often the data, $\Delta_i$, are susceptible
to a large source of noise, or a not-well-understood source of noise
that contaminates only one mode of the data.  For example, the average
value of $\Delta_i$ may be very poorly determined.  In this case, the
average is usually subtracted from $\Delta_i$.  Similarly, the monopole
and dipole are explicitly subtracted from the all-sky COBE/DMR data,
because the monopole is not determined by the data and the dipole is
local in origin.  In general, placing any constraint on the data or some
subset thereof, such as insisting that its average be zero, results in
additional correlations in $\Delta_i$.  We take this into account by
adding these additional correlations, $C_C$, to the noise matrix to
create a ``generalized noise matrix,'' $C_N$, where $C_N = C_n + C_C$.
In the limit that the amplitude of $C_C$ gets large, this is equivalent
to projecting out those modes which are now unconstrained by the data
\cite{THSVS}, but this scheme is numerically much simpler to implement.
Thus in the text below we always write the noise matrix as $C_N$ instead
of $C_n$.  The details of this procedure for handling the effect of
constraints are explained in Appendix \ref{app:snmode}.

Due to finite angular resolution and switching strategies
designed to minimize contributions from spurious signals
(such as from the atmosphere), the signal is generally
not simply the temperature of the sky in some direction,
$T(\hat x)$, but a linear combination of temperatures:
\be
s_i = \int d\Omega H(\hat x,\hat x_i) T(\hat x)
\ee
where $H(\hat x,\hat x_i)$ is sometimes called 
the ``beam map'', ``antenna pattern'' or ``synthesis
vector''.  If we discretize the temperature on the sky then we can
write the beam map in matrix form, $s_i = \sum_n H_{in} T_n$. 

The temperature on the sky, like any scalar field on
a sphere, can be decomposed into spherical harmonics
\be
T(\theta,\phi) = \sum_{\ell m} a_{\ell m}Y_{\ell m}(\theta,\phi).
\ee
If the anisotropy is {\it statistically}\/ isotropic, {\it i.e.},
there are no special directions in the mean, then the variance
of the multipole moments, $a_{\ell m}$, is independent of $m$ and
we can write:
\be
\langle a_{\ell m} a^*_{\ell'm'} \rangle = C_\ell \delta_{\ell\ell'}\delta_{mm'}.
\ee
For theories with statistically isotropic Gaussian initial
conditions, the angular power
spectrum, $C_\ell$, is the entire statistical content of the theory in
the sense that any possible predictions of the theory for the
temperature of the microwave sky can be derived from it
\footnote{Non-linear evolution will
produce non-Gaussianity from Gaussian initial conditions
but this is quite sub-dominant for $\ell \lesssim 1000$.}.
Even for non-Gaussian theories, the angular power
spectrum is a very important statistic, probably the most important one
for determining the viability of the most popular non-Gaussian theories.
However, the techniques we present in this paper for estimating the
power spectrum assume that the fluctuations in both the sky signal and
experimental noise are Gaussian.

The theoretical covariance matrix, $C_{Tii'}$, is related
to the angular power spectrum by
\be\label{eqn:CT}
C_{Tii'} = \sum_\ell {2\ell+1 \over 4\pi} C_\ell W_{ii'}(\ell) \, , 
\ee
where 
\be
\label{eqn:wl}
W_{ii'}(\ell) = \sum_{nn'}H_{in}H_{i'n'}P_\ell(\cos \theta_{nn'})
\ee
is called the window function of the observations and 
$\theta_{nn'}$ is the angular separation between the
points on the sphere labeled by $n$ and $n'$.  

Let us define the quantity 
${\cal C}_\ell \equiv \ell(\ell+1)C_\ell/(2\pi)$.  This is
useful for two reasons: 
it is the logarithmic average of ${\cal C}_\ell$ that
gives the variance of the data and (therefore)
for scale-invariant theories of structure formation,
${\cal C}_\ell$ is roughly constant at large scales.

Within the context of a model, the ${\cal C}_\ell$ depend on some
parameters, $a_p$, $p=1\ldots N_p$ which could be the Hubble
constant, baryon density, redshift of reionization, etc.  The
theoretical covariance matrix will depend on these parameters through
its dependence on ${\cal C}_\ell$.  We can now write down the likelihood
function for $a_p$, which is equal to the probability of the data
given $a_p$.
\begin{eqnarray}
  {\cal L}_\Delta(a_p)=&&P(\Delta|a_p)={1\over(2\pi)^{N/2}
    |C_T(a_p)+C_N|^{1/2}} \times\nonumber\\
 && \exp\left[-{1\over2}
    \Delta^T\left(C_T(a_p)+C_N\right)^{-1}\Delta\right].       
\end{eqnarray}
One can then search for the parameters $a_p$ that maximize
this likelihood.  

\subsection{Direct Evaluation of the Likelihood Function}

First, we must choose a set of parameters to characterize the
theoretical covariance, $C_T$.
For a given class of cosmological theories ({\it e.g.}, adiabatic 
perturbations
from inflation) we can calculate the power spectrum from some set of
parameters like the densities of various components, $\Omega_x$, the
shape of the primordial power spectrum, the Hubble constant, etc. A
detailed exploration of the cosmological parameter space constrained by
current CMB and large-scale structure data is given in\cite{BondJaffe}.
Alternately, we can describe the power spectrum by its actual value at
some discrete multipoles or bands of $\ell$.  Moreover, all of the
information in the experiment (again, for Gaussian theories) is captured
in the likelihood function for the power spectrum:
\begin{equation}
  P\left(\Delta|\{a_p\}\right)\propto 
  P\left(\Delta|\{C_\ell(a_p)\}\right)
\end{equation}
In this paper, we concentrate on the ${\cal C}_l$ parameterization in order to
determine the power spectrum directly from the data. In principle, we
would like to calculate the full likelihood as a function of the power
spectrum $P(\Delta|\{C_\ell\})$ for some $\ell\le\ell_{\rm max}$; at
the very least we would like to find the maximum of this $\ell_{\rm
  max}$-dimensional function, and its properties ({\it e.g.}, curvature or
``width'') around this maximum.

Searching such multi-dimensional spaces can be difficult; in this
case, each evaluation of the likelihood function is an expensive
$O(N^3)$ matrix manipulation and a brute force search through the
parameter space would take of order
$(\C_\ell/\delta \C_\ell)^{\ell_{\rm max}}$ such evaluations to reach an
accuracy of $\delta\C_\ell$.  
In our applications, we have found that the space is sufficiently
structureless that a simple iteration procedure works well for finding
the maximum. In addition, we do not use all of the individual $C_\ell$
values as separate parameters, since experiments do not have
uncorrelated information about bands of width
$\Delta\ell\lesssim 2\pi/\theta$, where $\theta$ is the angular extent
of the survey \cite{tegmarkdeltal}. 
For COBE/DMR, we bin in bands of width $\Delta\ell=2-3$ for
$\ell\ge25$ and only consider $\ell\le35$; above this multipole we
give the power spectrum a constant shape and amplitude (that of
COBE-normalized standard CDM, in this case). For
SK, we have tried bins of various widths, the choice of which we will
discuss below.

At the first iteration, we choose some appropriate starting ${\cal C}_\ell$.
For each $\ell$ (or band), we hold all other ${\cal C}_\ell$s fixed
while the one of interest is allowed to vary; in the appropriate
signal-to-noise basis, the likelihood as a function of this single
parameter is trivial to compute (see Appendix \ref{app:snmode}).  That
is, for each band labeled by $B$, we rewrite the correlation matrix as
\begin{equation}
        C_T+C_N= q_B C_B + C_{N^*}
\end{equation}
(no sum over $B$)
where the effective signal and noise matrices are given by
\begin{eqnarray}\label{eqn:cnstar}
        C_{Bii'} &=& \sum_{\ell\in{\rm B}} {2\ell+1\over4\pi}C_\ell W_{ii'}(\ell);\nonumber\\
        C_{N^*ii'} &=& C_{Nii'} + \sum_{L\not\in{\rm B}} {2L+1\over4\pi}C_L W_{ii'}(L).
\end{eqnarray}
and calculate the likelihood as a function of the adjustment factor
$q_B$ alone.  
After going through all the $\ell$ bands of interest, we then
update the starting power spectrum by multiplying the $C_\ell$s in
each band by the $q_B$ that maximized the likelihood function.
We then repeat.  Convergence is achieved when all the $q_B$s
equal unity.
For COBE/DMR, starting from COBE-normalized standard 
CDM (already a good fit) we achieved convergence at the few percent 
level after only two such iterations for $\ell\le20$; after 10
iterations, convergence is everywhere better than $10^{-4}$.

There is a drawback to the procedure as described so far, compared to
what could be achieved by more ambitious methods such as simulated
annealing \cite{numrec,knox95}.  Even though we find the maximum of the
likelihood function, we haven't accurately determined its shape---only
the shape along each ${\cal C}_\ell$ while the others are held constant
({\it i.e.}, parallel to the axes of the $\ell_{\rm max}$-dimensional
space).  And we have no estimate for the correlations between the
uncertainties in each estimate of ${\cal C}_\ell$.  Below, we shall see
how to use the Fisher matrix for an estimate of these correlations.
Clearly, a more ambitious minimization strategy would be preferable; we
have chosen not to implement one since the quadratic estimator to be
derived below achieves this end without any explicit likelihood
calculation. 

We have also considered the possibility of estimating each ${\cal
  C}_\ell$ assuming no other knowledge of all of the others. That is, we
have attempted to marginalize over the ${\cal C}_\ell$ values outside of
each band. This is equivalent to the procedure outlined in
Appendix~\ref{app:snmode} for marginalizing over removed constraints
(averages, dipoles, etc.)\/ and foreground templates. However, in this
case, the method fails to constrain the power spectrum. In performing
this marginalization, we effectively allow an arbitrary amount of noise
consistent with {\em any power spectrum at all}\/ outside of the band of
interest. That is, we multiply the second term in Eq.~\ref{eqn:cnstar}
by a very large number to make the variance in those modes larger than
the noise or (expected) signal.  For a perfect, all-sky observation,
this would not be a hindrance since all the multipoles are independent.
For any realistic observation, however, there is aliasing of different
multipoles together; some modes of the data (defined, for example, by
the eigenmodes of Appendix \ref{app:snmode}) that are being
marginalized over will have nonzero contributions from within the $\ell$-band
of interest. Thus, the new {\em noise}\/ spectrum alone will span the
space of possible {\em signals}, consistent with having no power at all
in the band. This just reinforces the idea that any unknown noise in the
observation should ideally be completely ``orthogonal'' to the
quantities we are attempting to estimate (which will often be the case
when the marginalization technique is used for experimental constraints
or foreground removal).

\subsection{Gaussian Approximation to the Likelihood Function}

If the likelihood function is continous and has a peak then
it can be approximated as a Gaussian near the peak.  
For well-constrained parameters this approximation
should be good except in the tails of the distribution.  
A Gaussian approximation to the likelihood function can be obtained by
truncating the Taylor expansion of $\ln {\cal L}$ about $a_p$ at
second order in $\delta a_p$: \bea
\label{eqn:Taylor}
\ln {\cal L}(a+\delta a) &=& \ln {\cal L}(a) + \sum_p
{\partial \ln{\cal L}(a) \over \partial a_p} \delta a_p
\nonumber\\
&+& {1\over 2}\sum_{pp'}{\partial^2 \ln{\cal L}(a) \over \partial a_p 
\partial a_{p'}}\delta a_p \delta a_{p'}.
\eea
This Gaussian approximation is useful because now,
instead of making multiple evaluations of the likelihood
function,  we can directly solve for the $\delta a_p$ that maximize
it:
\be
\label{eqn:lnL}
\delta a_p = -\sum_{p'}\left[{\partial^2 \ln {\cal L}(a)
\over \partial a_p 
\partial a_{p'}} \right]^{-1} {\partial \ln {\cal L}(a) \over
\partial a_{p'}}.
\ee
The first derivative is given by:
\bea
{\partial \ln {\cal L}(a) \over \partial a_{p'}} =
{1\over 2}{\rm Tr}\left[ \left(\Delta \Delta^T - 
C\right)\left(C^{-1}C_{T,p'}C^{-1}\right)\right]
\eea
and the second derivative by
\bea
\label{eqn:curvature}
{\cal F}_{pp'}^{(a)} & \equiv & 
-{\partial^2 \ln {\cal L}(a)\over \partial 
a_p \partial a_{p'}} \nonumber\\
&=& {\rm Tr}\big[ \left(\Delta \Delta^T - 
C\right)(C^{-1}C_{T,p}C^{-1}C_{T,p'}C^{-1}\nonumber\\
& & \phantom{{\rm Tr}\big[ \left(\Delta \Delta^T - 
C\right)(C^{-1}C_{T,p}}
-{1\over 2}C^{-1}C_{T,pp'}C^{-1})\big]\nonumber\\
&+&{1\over 2} {\rm Tr}\left(C^{-1} C_{T,p}C^{-1}C_{T,p'}\right)
\eea
where Tr is the trace, 
$C \equiv C_T + C_N$ is the total covariance matrix 
and $_{,p} \equiv \partial/\partial a_p$.
We call the second derivative the curvature matrix and
give it the symbol ${\cal F}^{(a)}$ where the $(a)$
indicates that we have taken the derivative of $\ln{\cal L}$ 
with respect to $a$.  

To the extent that the likelihood function is not Gaussian, we will
not have correctly solved for its maximum.  Thus we iterate.  The
closer we get to the maximum, the better the quadratic approximation
to $\ln {\cal L}$ will become.  This is exactly the
Newton-Raphson method for finding the zero of $\partial \ln {\cal L}/
\partial a_p$.  The procedure is not fool-proof---there is
the risk of getting trapped in a local extremum.  In practice
we have found the likelihood function to be sufficiently structureless
that this is not a problem.  

\subsection{Quadratic Estimator}

The above procedure is not exactly what we do in practice.
Calculating the curvature matrix is a 
computationally intensive procedure.  Matters simplify significantly if
we settle for the ensemble average quantity, called the Fisher matrix,
$F$:
\bea\label{eqn:fish}
F^{(a)}_{pp'} & \equiv & 
\langle {\cal F}^{(a)}_{pp'} \rangle \nonumber\\
&=& {1\over 2}{\rm Tr}\left( C^{-1} C_{T,p}C^{-1}C_{T,p'}\right).
\eea 
When taking this ensemble average, denoted by 
$\langle ... \rangle$, we assume that the theory is correct and
therefore that $\langle \Delta \Delta^T \rangle = C$.  

Note that the Fisher matrix, like the curvature matrix,  
is defined with respect to particular
parameter choices.  If we transform to a new set of
parameters, $\tilde a_p$ then the Fisher
matrix for these new parameters is $F^{(\tilde a)}= 
Z^{-1} F^{(a)} (Z^{-1})^T$, where $Z_{pp'} = \partial \tilde a_p /
\partial a_{p'}$.  Tegmark offers a proof of this
\cite{tegmarkoptimal}; with our approach it is obvious from the
definition of the curvature matrix in Eq. \ref{eqn:curvature}.

Replacing the curvature matrix with the Fisher matrix makes our 
estimator for $a_p$ quadratic in the data, $\Delta$:
\be
\label{eqn:quadest}
\delta a_p = {1\over 2}\sum_{p'}(F^{(a)})^{-1}_{pp'}
{\rm Tr}\left[\left( \Delta \Delta^T-C\right)\left( 
  C^{-1} C_{T,p'} C^{-1}\right)\right].
\ee
This is what we call the quadratic estimator.  The right hand-side depends
on $a_p$, so we pick an initial $a_p$,
calculate the correction $\delta a_p$, and then repeat for the
new value of $a_p$.  Note that the power spectrum estimate is not
constrained to be positive-definite---a point we discuss below.

If we assume that the input theory is correct, then 
$\langle \Delta \Delta^T \rangle = C$ and therefore
Eq.~\ref{eqn:quadest} implies 
$\langle \delta a_p \rangle=0$.  Similarly, one can work
out that $\langle \delta a_p \delta a_{p'} \rangle = 
(F^{(a)})^{-1}_{pp'}$.  This is to be expected since 
for a Gaussian distribution, the two-point function is the
inverse of the curvature matrix.

Although the quadratic involves using the Fisher matrix $F$ as an
approximation to the full curvature matrix ${\cal F}$, both procedures
iterate to the {\em same}\/ parameters, the maximum of the likelihood
function. This is because both ${\cal F}$ and $F$ are invertible, 
so $\delta a_p=0$ from either procedure implies $\partial\ln{\cal
  L}/\partial a_p=0$. Thus, when applied iteratively, {\em the quadratic
  estimator will find the exact location of the likelihood peak}; the
only approximation comes in using the Fisher matrix to approximate the
errors, rather than the full curvature matrix (and below we show that in
the cases studied, this is a very good approximation; moreover, having
found the location of the peak, the curvature there can be calculated
explicitly if necessary).  

Our procedure is very similar to that of the Levenberg-Marquardt
method \cite{numrec} for minimizing a $\chi^2$ with
non-linear parameter dependence.  There the curvature matrix
(second derivative of the $\chi^2$) is replaced by its
expectation value and then scaled according to whether the
$\chi^2$ is reduced or increased from the previous iteration.  
Similar manipulations of the Fisher matrix 
may possibly speed convergence of the likelihood maximization,
although one would want to do this without direct evaluation of the
likelihood function.

In our applications to COBE/DMR and SK we have found that 
iteration converges quickly.  Iteration is especially important
for the calculation of the error covariance matrix.  Without iteration,
the errors are determined entirely by the initial theoretical
assumptions and are not influenced by the data.  (Of course,
this is exactly why the Fisher matrix has been so useful in
determining how well future observations will be able to determine
parameters.)

As we have defined it so far, the quadratic estimator with the
iteration procedure is a method for finding the maximum of the
likelihood.  Only if one takes the prior probability to be uniform
in the parameters is this equivalent to maximizing the posterior
probability.  We
could, of course, include different priors directly in the definition
of the estimator.  The derivation would then begin by changing
Eq.~\ref{eqn:Taylor} to 
a Taylor expansion of $\ln{P_{\rm post}}$ where 
$P_{\rm post} \propto {\cal L}P_{\rm prior}$ is the posterior
probability distribution and $P_{\rm prior}$ is the (differentiable) prior
distribution.

To see how the quadratic estimator works, we can take
a one-dimensional example.  Consider a function $f$, that
is approximately quadratic.  If we take its first and second
derivatives about some point, $x_0 \ (= 0.7$ in the figure),
we can construct the function $f_Q$ which approximates $f$.  By finding the value
of $x$ that maximizes $f_Q$ we have a guess as to
the maximum of $f$.  Now, for a further refinement of the 
estimate, a new $f_Q$ can be calculated
based upon the properties of $f$ at this new value of $x$. (Note that
the full quadratic estimator of Eq.~\ref{eqn:quadest} includes the further
approximation of using the Fisher matrix (Eq.~\ref{eqn:fish}) rather than the
actual curvature matrix (Eq.~\ref{eqn:curvature}) for the second
derivative of the log-likelihood.) 

\begin{figure}[bthp]
\plotonecita{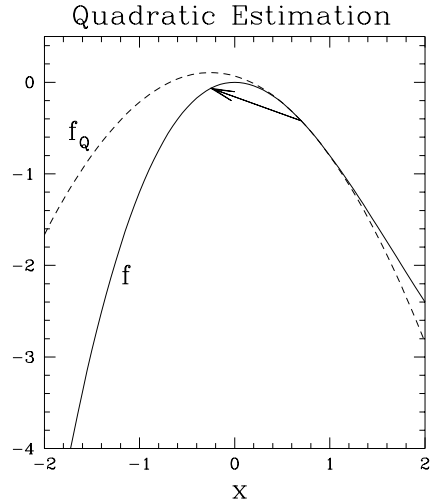}
\plotonecfpa{demquad.eps}
\caption[One-dimensional quadratic estimate]{\baselineskip=10pt 
A one-dimensional example of quadratic estimation.}
\label{fig:one-D}
\end{figure}

The applications we discuss in the following all use the ${\cal C}_\ell$s
as the parameters $a_p$.   In this case, 
\begin{equation}\label{eqn:CTell}
  C_{Tii',\ell}\equiv{\partial C_T\over\partial{\cal C}_\ell}=
  {{\ell+1/2}\over\ell(\ell+1)} W_{ii'}(\ell)\simeq W_{ii'}(\ell)/\ell.
\end{equation}
We also consider the power spectrum averaged over some bands $B$ with some
assumed shape ${\cal C}_\ell^{\rm shape}$; in that case, we average the
above weighted by the shape:
\begin{equation}\label{eqn:CTB}
  C_{Tii',B} = \sum_{\ell\in B} C_{Tii',\ell}{\cal C}_\ell^{\rm shape}.
\end{equation}
However, there is also
the interesting possibility of taking the $a_p$ as
the cosmological parameters that affect the spectrum,
$\Omega$, $h$, $n_S$, $\Omega_b$, etc.  Iteration in
this case should also converge to the likelihood maximum.

We note that the quadratic estimator discussed here can also be
derived by finding the quadratic function of the data that is
unbiased and has minimum variance.  For a
full discussion of the quadratic in this context,
see \cite{tegmarkoptimal,Hamilton,TegHam}.   The quadratic
function of the data derived this way is the same as 
Eq.~\ref{eqn:quadest}.  However, the estimate is only unbiased
if there is no iteration.  Since the end point of (successful)
iteration is the maximum likelihood, the iterated estimator is,
like all maximum likelihood estimators, only asymptotically
unbiased. 

The methods we have used can also be applied to optimal determination
of the correlation function in angular bins. The optimal signal plus
noise weighting suggested for correlation function determination
differs from the usual ${\rm diag}[C_n^{-1}]$ weighting applied to
COBE/DMR.

\subsection{Single Bandpower Estimation}   

It has now become conventional to characterize switching experiments
which covered small patches of the sky by a single
bandpower\cite{Bond}, placing the estimated power at a location
related to the window function of the experiment. In this case, there
is just one parameter to determine. The quadratic statistic reduces to 
\begin{eqnarray}
&& Q_{B}= {\Delta^\dagger C^{-1} C_T C^{-1} \Delta - \Tr C_NC^{-1} C_T
C^{-1} \over  \Tr C_TC^{-1} C_T C^{-1} } \, . \label{eqn:BCstat}
\end{eqnarray}
If the optimal weight $C^{-1}$ is replaced by the diagonal part of
$C_n^{-1}$, then this is related to the quadratic statistic proposed by
Boughn and Cottingham\cite{BCstat}, which has been applied to the
COBE/DMR and FIRS data using Monte Carlo simulations to define its
distribution. With the optimal weighting and the proper inclusion of
constraints in $C_N$, the values of $Q_{B}$ and its error estimation
are of direct use. As discussed above, the iterated quadratic estimator
for the amplitude will converge to the maximum likelihood value.
 The parameter $Q_B$ could be any squared amplitude
characterizing the assumed theoretical ${\cal C}_\ell$, such as the
$\sigma_8^2$ used to characterize the strength of the power spectrum
on cluster scales. To translate to an average bandpower one must
evaluate $Q_B \avrg{{\cal C^{\rm shape}}_\ell}_B$, 
using an appropriately weighted
average of ${\cal C^{\rm shape}}_\ell$ 
over the single band $B$. Issues associated
with such averaging are addressed in \S~\ref{sec:binning}. Current and
future experiments cover large enough patches of the sky that
characterizing their results by single bandpowers is not useful, but
evaluation of power spectrum normalization amplitudes (such as
$\sigma_8$) for particular theories will always be of use.

\section{Application to COBE/DMR}
\label{sec:cobedmr}

We first apply these methods to the anisotropy measurements of the
COBE/DMR instrument\cite{DMR,gorskiCl}. The DMR instrument actually
measured a complicated set of temperature differences $60^\circ$ apart
on the sky, but the data were reported in the much simpler form of a
temperature map, along with appropriate errors (which we have expanded
to take into account correlations generated by the differencing
strategy, as treated in \cite{Bond94}, following \cite{LineSmoot}). 
The calculation of the theoretical correlation matrix includes 
the effects of the beam, digitization of the time stream, 
and an isotropized treatment of pixelization, using the
table given by Kneissl and Smoot \cite{GneSmoot}, 
modified for resolution 5.  We use a weighted
combination of the 31, 53 and 90~GHz maps. Because most of the
information in the data is at large angular scales, we use the maps
degraded to ``resolution 5'' which has 1536 pixels. Further, we cannot
of course observe the entire CMB sky; we use the most recent galactic
cut suggested by the DMR team\cite{DMR}, leaving us with 999 pixels to
analyze.  We use the galactic, as opposed to ecliptic, pixelization.

For both methods we iterated 28 parameters: ${\cal C}_2$ to ${\cal
C}_{24}$ individually, ${\cal C}_{25}$ to ${\cal C}_{32}$ grouped into
bins of width 2 and finally ${\cal C}_{33}$ through ${\cal C}_{35}$
grouped into one bin.  Binning is described in more detail prior to
the Saskatoon application where it is much more important.

\begin{figure}[bthp]
\plotonecita{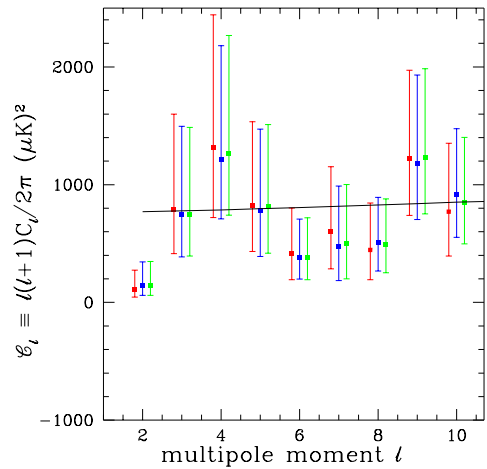}
\plotonecfpa{iterate_errbar_lindmr0to3.eps}
\caption[linear iteration of DMR]{\baselineskip=10pt 
  Maximum-likelihood power spectra from iterative direct evaluation of
  the likelihood function.  The curve is the zeroth iteration:
  COBE-normalized standard CDM.  The points with error bars are, from
  left to right, the results of the first to third iterations. Here, we
  define the error bars by a likelihood ratio of $e^{-1/2}$ from the peak.}
\label{fig:lindmr}
\end{figure}

\begin{figure}[bthp]
\plotonecita{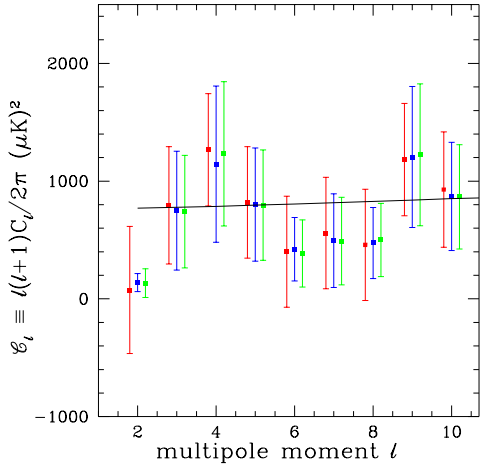}
\plotonecfpa{iterate_errbar_quaddmr0to3.eps}
\caption[Quadratic iteration of DMR]{\baselineskip=10pt 
Iterative quadratic estimation. The curve is the zeroth iteration: 
COBE-normalized standard
CDM.  The points with error bars are, from left to right,
the results of the first to third iterations.}
\label{fig:quaddmr}
\end{figure}

\begin{figure}[bthp]
\plotonecita{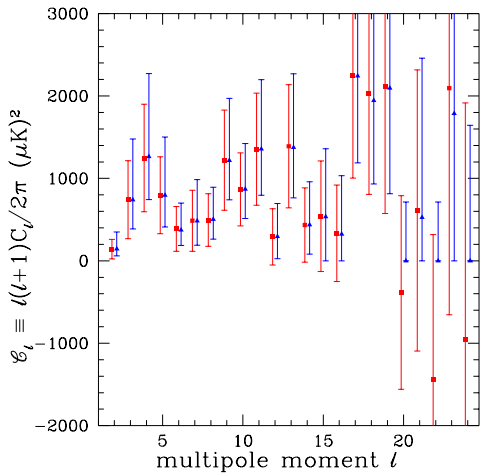}
\plotonecfpa{quadlindmr.eps}
\caption[quadratic/linear comparison]{\baselineskip=10pt 
  We compare the results of the quadratic and direct evaluation
  iteration schemes. At each $\ell$, the left error bar (square symbol)
  is for the quadratic, the right (triangle) is for the direct evaluation.}
\label{fig:quadlindmr}
\end{figure}

In Figures \ref{fig:lindmr} and \ref{fig:quaddmr} 
we see the results of the iterative procedures
described in the previous section.  Figure~\ref{fig:lindmr} 
shows the results
of direct evaluation and  Fig.~\ref{fig:quaddmr} 
shows the results
of quadratic estimation.  Moments $\ell > 10$ are
not shown to avoid clutter.  From left to right are the first
to third iterations, together with their error bars.  The solid
line is the starting point we chose, the power spectrum for
COBE-normalized standard CDM. For this method, we define
the estimated ${\cal C}_\ell$ as the maximum of the likelihood function,
and the errors by the value of ${\cal C}_\ell$ where the likelihood
drops by a factor $e^{-1/2}$ from that maximum.

First we will discuss the direct evaluation method. The iteration
converges rapidly.  The maximum likelihood values of a fourth
iteration (shown in Fig.~\ref{fig:quadlindmr}) typically differ 
from the third by 1--3\% of the
error bars (for $2\le\ell \le 19$)  with a maximum deviation of 7\% at
$\ell=12$. In the limit that the moments were independent, there would
be no need for iteration; iteration is only necessary because of the
influence the value of one band has on the best value of another.  The
rapidity of the convergence is expected because, as we will see below,
the moments are in fact fairly uncorrelated.
We remind the reader that
the error bars given by this method---indeed the whole probability
distribution for each ${\cal C}_\ell$---are calculated by holding the
others fixed.

Iteration is also quite rapid for the quadratic estimator: the maximum
likelihood values of a fourth iteration (shown in
Fig.~\ref{fig:quadlindmr}) differ from the third by better than 1\% of
the square root of the variance for $\ell\leq 24$, except for the
quadrupole and $\ell=20$ which are slightly worse, converging to 3\%.
Just like the direct method, most of the change in the maximum
likelihood estimate occurs in the first iteration.

Unlike the direct method, the error bars of the first iteration are
quite different from the error bars of the later iterations.  That is
because the error bars (the Fisher matrix) do not depend on the data,
but only on the input power spectrum.  Therefore the data have had no
effect on the error bars until the second iteration is reached.  To the
extent that the distribution is Gaussian, these error bars accurately
represent the uncertainty on each parameter; they take into account the
correlations with the other parameters.  The largest changes in the
error bars from 1st to 2nd, 2nd to 3rd and 3rd to 4th are 610\%
($\ell=2$), 60\% ($\ell=2$) and 6.5\% ($\ell = 6$), respectively.  From
the 3rd to the 4th, most of the changes are less than 1\%.

In the previous section it was claimed that the curvature matrix is a
good approximation to the Fisher matrix.  We have explicitly checked
this for the final iteration and find that for $\ell < 20$ most of the
Fisher matrix and curvature matrix derived error bars agree with each
other to better than 4\%.  The worst cases are $\ell=4$ and $\ell=5$ at
13\% and 15\%.

\begin{figure}[bthp]
\plotonecita{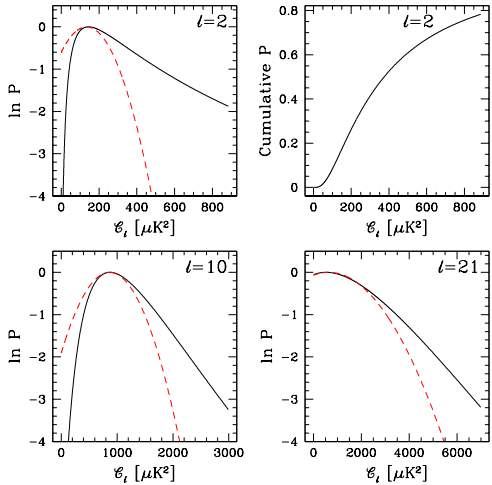}
\plotonecfpa{likefigs.eps}
\caption[Probability distributions]{\baselineskip=10pt 
Probability Distributions for individual ${\cal C}_\ell$ values, as
labeled, for a prior uniform in $\C_\ell$.
The solid curve is the true likelihood from the last iteration
of the full evaluation; the dotted curve is the Gaussian approximation
from the last iteration of the quadratic procedure. For $\ell=2$, we
also show the cumulative probability distribution, properly normalized
to unit probability as ${\cal C}_\ell\to\infty$.}
\label{fig:P(Cl)}
\end{figure}

Not only do these methods converge, but they converge to the same
power spectrum, as we see in Fig.~\ref{fig:quadlindmr}.  The
differences between the final iterations are less than 2\% of the
quadratic error bars for $\ell<20$, except for a 4\% difference at
$\ell=18$; at higher moments, the methods often do not detect positive
power.  Note that at multipoles where both methods do detect nonzero
power, the quadratic method gives error bars which are systematically
smaller (than those of the direct method) in the direction of positive
power, and systematically larger towards lower power. This can be
understood as a result of the considerable non-Gaussian skewness of
the distribution of power, as seen in Fig.~\ref{fig:P(Cl)}.
Also note that when the likelihood maximum is at zero power, the
quadratic estimate is at (physically meaningless) negative power.  This
is to be expected since the existence of a maximum at $\C_\ell = 0$
implies $\partial \ln{\cal L} / \partial \C_\ell \leq 0$, and therefore
the Gaussian fit to $\ln{\cal L}$ at $\C_\ell = 0$ will peak at $\C_\ell
\leq 0$.

We have also checked that using the full resolution 6 data (3881
pixels after the galactic cut) changes the results of the
maximum-likelihood estimate for the power spectrum by much less than one
sigma.  We have checked in detail using the direct evaluation, for which
the resolution 6 results differ from those at resolution 5 by less than
5\% for $\ell\le15$, except at $\ell=6$--$9$ where the difference is
almost 10--20\% and at $\ell=12$ and $\ell=14$, where the difference is
nearly 50\%, still smaller than the large error at these $\ell$; the
higher resolution data give an overall normalization that differs by 4\%
(compared with an error of 14\%) from that of the best quadratic
computed at resolution 5. These differences are consistent with those
observed for different pixelizations and galactic
cuts\cite{DMR,gorskiCl}; note that both the direct evaluation and
quadratic procedures converge with considerably higher precision than
these intrinsic errors, even for $\ell\gtrsim15$ where the pixelization
differences become important and, simultaneuously, the noise begins to
dominate.

\begin{figure}[htbp]
\plotonecita{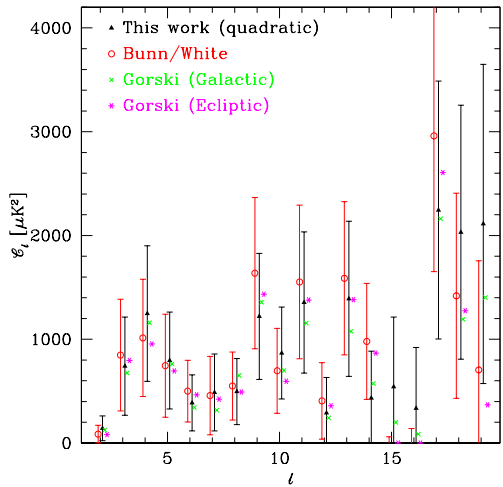}
\plotonecfpa{bwcompare_b2.eps}
    \caption{Comparison of different groups' power spectrum estimates,
      as marked. 
      Gorski computes power spectra in both ecliptic and galactic
      pixelizations of the sky.}
    \label{fig:compare}
\end{figure}

We also agree at least qualitatively with other calculations that we
have compared to, in all cases (with detected power) well within the
various reported error bars.  In Fig.~\ref{fig:compare} we show a
comparison of our quadratic results with those of
\cite{BunnWhite,gorskiCl}, both of whom use a maximum likelihood method.
Gorski\cite{gorskiCl} uses a complete search through parameter space
with ``cut-sky spherical harmonics'' to speed up the calculation; Bunn
\& White \cite{BunnWhite} also use the Signal-to-Noise transformation of
Appendix \ref{app:snmode} to increase speed.  The results of our first
quadratic iteration also have qualitative agreement with Tegmark's
implementation of the quadratic estimator\cite{tegmarkoptimal}.

The fact that three completely different methods achieve similar results
lends support to the claim that the final estimates are unaffected by
the choice of initial starting place, and the stronger claim that they
would have resulted from {\it any}\/ starting place.  From the Fisher
matrix and from the probability distributions of Fig.~\ref{fig:P(Cl)} it
should be evident that this likelihood space is fairly structureless.
We could have started anywhere and converged to the same place, although
perhaps slightly less rapidly.  We note though that if the correlations
were stronger between the different ${\cal C}_\ell$s, the direct method
would be less robust.  In particular, if the initial power spectrum were
much too large, then each multipole moment would try to make up for this
all by itself by coming out very small.  Thus there could be large
oscillations---conceivably without convergence.  In addition, these
correlations, combined with the width of the likelihood function, imply
that our iterative direct evaluation method for finding the peak may not
converge to a unique maximum, as values oscillate between iterations; in
practice, we have found that the changes remain much smaller than the
size of the error bars, as noted above.  Such a broad likelihood
function indicates that the data do not strongly prefer a unique
maximum. Nonetheless, if we desire to find the exact location of the
peak, a more complete search through the many-parameter space (as in
\cite{BunnWhite,gorskiCl}) or the use of the quadratic method will be
necessary.
 
The probability distributions of the parameters are different for the
two different methods because of the approximation of independence by
the direct method and the approximation of Gaussianity by the quadratic
method.  We can see those differences in Fig.~\ref{fig:P(Cl)}.  The
departure from Gaussianity is most dramatic for the quadrupole.
According to the Gaussian distribution of ${\cal C}_2$, COBE-normalized
CDM with $\C_2 = 770\ \muK^2$ is over five standard deviations away from 
the mean, highly ruled
out.  But the strong skewness of the exact likelihood function has 25\%
of the probability for ${\cal C}_2$ above $770\ \muK^2$.  This is more
probability than there is above only $1 \sigma$ for a Gaussian
distribution!\footnote{Also these quadrupole probability distributions
  do not take into account the possibility of foreground contamination.
  The DMR team\cite{dmrFGs} have carefully analyzed the foreground
  contamination and report ${\cal C}_2=(273\pm185\pm360)\ \muK^2$ with
  statistical and systematic errors.}
As $\ell$ increases the distributions become more Gaussian.  The
distribution for $\ell = 21$ is well-approximated by a Gaussian as
expected from the central limit theorem since there are approximately 30
independent modes of roughly equal weight contributing to the
constraint.

The highly non-Gaussian nature of some of these distributions implies
that other definitions of the point estimation and the error bars are
possible. First, we could consider the mean or median of the
distribution, rather than its maximum, and define errors by the amount
of enclosed probability. Second, we could also have used different prior
probabilities for the ${\cal C}_\ell$. Throughout the paper, we use a
prior uniform in ${\cal C}_\ell$, equivalent to equating the posterior
distribution with the likelihood itself. When the data constrain the
power strongly ({\it i.e.}, small error bars), the result is insensitive
to the choice of the prior; in other regimes, such as the quadrupole,
${\cal C}_2$, the prior has more significance. To investigate this, we
have also tried other possible prior distributions, along with the
definition of the point estimate by the median of the distribution. A
prior ${\rm P}({\cal C}_\ell)d{\cal C}_\ell\propto d{\cal
  C}_\ell/\sqrt{{\cal C}_\ell}$ (which is equivalent to a prior uniform
in $\sigma_{\rm th}=({\cal C}_\ell)^{1/2}$) gives a median ${\cal C}_2$
60\% higher than the likelihood maximum; the highly skewed distribution
means that for a constant prior the median is 166\% higher, while a
prior uniform in $\ln{\cal C}_2$ has a median only 5\% higher. Finally,
we have also tried a ``Fisher Prior,'' which uses the element of the
Fisher matrix (Eq.~\ref{eqn:fish}) corresponding to $a_p=\sigma_{\rm
  th}$ to determine the expected amplitude,
\begin{equation}
  P(\sigma_{\rm th}^2) \propto F_{\sigma\sigma}^{1/2}\propto 
  \left({\rm Tr}\left[
    {\partial\ln\left(C_N+\sigma_{\rm th}^2 C_T\right)\over\partial\sigma_{\rm
    th}^2}\right]^2\right)^{1/2}
\end{equation}
which is uniform in ${\cal C}_\ell\propto\sigma_{\rm th}^2$
at low amplitudes, but uniform in $\ln{\cal C}_\ell$ at high amplitudes,
where the smooth transition is determined by the scale at which
signal-to-noise becomes about one. For this prior, the median is about
20\% higher than the maximum likelihood.

\begin{figure}[bthp]
  \plotonecita{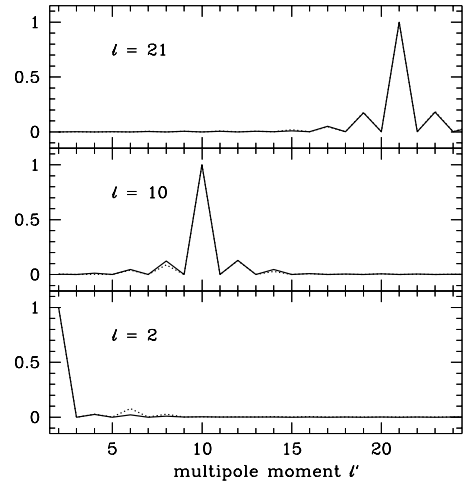}
  \plotonecfpa{dmrfishrows_2_10_21.eps}
\caption[Fisher matrix image]{\baselineskip=10pt 
  Rows of the normalized DMR Fisher matrix (see text), 
  at $\ell=2,10,21$. The solid lines show the matrix at the
  zeroth iteration; the dashed lines for the final iteration.}
\label{fig:fishdmr}
\end{figure}

In Fig.~\ref{fig:fishdmr} we show the normalized Fisher matrix,
$F^{(\C)}_{\ell \ell'}/\sqrt{F^{(\C)}_{\ell
\ell}F^{(\C)}_{\ell'\ell'}}$ to indicate the level of correlations
between the different $\C_\ell$s.  The off-diagonal terms are due to
the inhomogeneous coverage, the most drastic component of which is due
to the galactic cut.  This cut discards all map pixels with
galactic latitude $|b| \leq 20^\circ$, with some modifications
motivated by the DIRBE dust map\cite{DMR}.  A map with a $|b| \leq
20^\circ$ cut and otherwise homogeneous coverage would result in zero
overlap between $Y_{\ell m}$s with opposite parity which explains the
near zero values of the Fisher matrix for $\ell'-\ell$ odd
\cite{BunnWhite}.  Modes with similar parity do mix and hence the
non-zero elements at $\ell' = \ell \pm 2$.  Even these off-diagonal
terms though are much smaller than the diagonal, especially for the
lower multipole moments which are determined by modes with higher
signal-to-noise.  Iteration does not have much effect on the
normalized Fisher matrix; the off-diagonal components are largely a
result of the coverage geometry.

\section{Methods: Binning and Rebinning}
\label{sec:binning}

For the same reason that limited extent in the time domain leads 
to limited spectral resolution in the frequency domain, uncertainties
in ${\cal C}_\ell$ and ${\cal C}_{\ell'}$ are strongly correlated when
$\ell \lesssim 2\pi/\theta$ where $\theta$ is the linear extent of
the observed region \cite{tegmarkdeltal}.  
Thus binning moments together in bins
of width $\Delta \ell \simeq \pi/\theta$ is a sensible thing to
do.  Because of the experimental noise, final bins may need to
be even coarser to prevent the error bars from being excessively
large.

We view binning as a two-step procedure: an initial fine binning
followed by a rebinning to coarser bins.  The reason for the first
step is that we want to know, within each coarser bin, where the
constraining information is.  
The finer binning gives us this knowledge.  For
pedagogical reasons, we start with a discussion of rebinning and then
discuss the initial binning.

\subsection{Rebinning} \label{subsec:rebin}

We assume here that the initial binning is the finest possible,
$\Delta\ell = 1$, since this makes for the simplest exposition.  
It is easily generalized to arbitrary initial binning.
For reasons that will become clear later, we begin our discussion of 
this rebinning procedure by reparameterizing the power
spectrum in terms of an assumed spectral shape,
${\cal C}_\ell^{\rm shape}$.  Thus the parameters
we are trying to estimate are no longer ${\cal C}_\ell$ directly,
but the deviation from the assumed shape, given by $q_\ell$:
\be
{\cal C}_\ell = q_\ell {\cal C}_\ell^{\rm shape}.
\ee
If our estimates of individual $q_\ell$ are too noisy then
we can average them together into coarser bins, which
we will label by the subscript $B$.  We
wish to do this in a ``minimum variance'' manner.  That is,
we want to find $Q_B$ that minimizes
\be
\chi^2 = \sum_{\ell\ell'}\left(Q_B-q_\ell\right)F^{(q)}_{\ell\ell'}
\left(Q_B-q_{\ell'}\right)
\ee
where the sum (like all sums in this subsection) extends 
over the width of the new and coarser bin.
The Fisher matrix appears here because, in the Gaussian approximation
to the likelihood function, the Fisher matrix is the inverse of the
parameter covariance matrix.  Complications due to non-Gaussianity
are discussed in \S~\ref{sec:RadComp}.

It is easy to show that the solution to this minimization problem
is given by
\be
\label{eqn:QB}
Q_B = {\sum_{\ell\ell'} q_\ell F_{\ell\ell'}^{(q)} \over \sum_{\ell\ell'} F_{\ell\ell'}^{(q)}}.
\ee
The new parameters $Q_B$ have the Fisher matrix, $F^{(Q)}_{BB'} =
\sum_{\ell\ell'}F^{(q)}_{\ell\ell'}$ 
where the sum over $\ell$ extends across bin
$B$ and the sum over $\ell'$ extends across bin $B'$.
We see that $Q_B$ averages $q_\ell$ over the filter
$f^{(q)}_{B\ell} = \sum_{\ell'} F^{(q)}_{\ell \ell'}$.  
The $^(q)$ superscript indicates that this filter is for
averaging $q_\ell$s. 

As the constraints on the power spectrum become tighter, it
is inevitable that we will move from plotting averages of $\C_\ell$
(band-powers) to plotting $q_\ell$ in what we call $q$-space, 
or deviation space.  We show
some examples of this later in \S~\ref{sec:forecast} where we simulate future
data sets.   Therefore it is worth exploring this space a little
further.  One question to answer is: what $\ell$ value should be
used for locating $Q_B$ horizontally on a graph?  We advocate
choosing this $\ell_{\rm eff}$ so that for 
a band ranging from $\ell_1$ to $\ell_2$
\be
\sum_{\ell = \ell_1}^{\ell_{\rm eff}}f^{(q)}_{B\ell} = 
\sum_{\ell = \ell_{\rm eff}}^{\ell_2}f^{(q)}_{B\ell}.
\ee
With this definition, 50\% of the weight that constrains $q_B$
comes from $\ell_1 < \ell < \ell_{\rm eff}$ and the other 50\% comes
from $\ell_{\rm eff} < \ell < \ell_2$.  

Although comparison of theories with the data will occur in $q-$space,
we wish to translate our values into the familiar $\C_\ell$-space.  To
do this we must define a suitable average of $\C_\ell^{\rm shape}$
over bin $B$, $\C_B^{\rm shape}$, with which to multiply $Q_B$ and a suitable
$\ell$ value at which to plot the error bar, $\ell_{\rm eff}$. The best
weighting to use for this is debatable.  We emphasize that 
the ambiguities associated with the translation from $Q_B$ to a power
estimate, 
$\C_B$ only affect plotting---not the comparison of
theory with data.  Furthermore, we have tried several different
weighting schemes and found negligible differences in their
values of $\ell_{\rm eff}$ and $\C_B$, so long as they are proportional
to $f_{B\ell}^{(q)}$ which encodes the signal-to-noise information
in the band.  

To motivate a particular averaging we first
rewrite Eq.~\ref{eqn:QB} in terms of ${\cal C}_\ell$ and its Fisher
matrix:
\be
\label{eqn:Cl2QB}
Q_B = {\sum_{\ell\ell'} {\cal C}_\ell F_{\ell\ell'}^{({\cal C})} {\cal C}_{\ell'}^{\rm shape}
\over \sum_{\ell\ell'} {\cal C}_\ell^{\rm shape} F_{\ell\ell'}^{({\cal C})} 
{\cal C}_{\ell'}^{\rm shape}}.
\ee
The relation between $Q_B$ and $\C_\ell$ in the above equation suggests
that the following filter be used to calculate $\C_B^{\rm shape}$:
\be
\label{eqn:fBl}
f^{(\C)}_{B\ell} = \sum_{\ell'} F_{\ell\ell'}^{(\C)}{\cal
  C}_{\ell'}^{\rm shape}=f^{(q)}_{B\ell}/\C^{\rm shape}_\ell
\ee
since this is the weighting of each ${\cal C}_\ell$ in Eq.~\ref{eqn:Cl2QB}.
Therefore to make our power estimates we use
\be
\label{eqn:aveshape}
{\cal C}_B^{\rm shape} = {\sum_\ell f^{(\C)}_{B\ell} {\cal C}_\ell^{\rm shape} 
\over \sum_\ell f^{(\C)}_{B\ell}}.
\ee
with the result that
\be
{\cal C}_B = Q_B{\cal C}_B^{\rm shape} = {\sum_\ell f^{(\C)}_{B\ell} {\cal C}_\ell 
\over \sum_\ell f^{(\C)}_{B\ell}}.
\ee
The role of the filter function, $f^{(\C)}_{B\ell}$ 
is exactly that of 
$W_\ell/\ell$ in the band-power procedure of \cite{Bond}, 
where $W_\ell$ is the trace of the window function matrix defined in 
Eq.~\ref{eqn:wl}.
We will develop this connection more later.  For now, we define
$\ell_{\rm eff}$, $\ell^+$ and $\ell^-$, exactly as was done in
\cite{Bond}, so that we can plot data
points properly located in $\ell$ space with horizontal error bars:
\be
\ell_{\rm eff} = {\sum_\ell \ell f^{(\C)}_{B\ell}\over\sum_\ell f^{(\C)}_{B\ell}}
\ee
and $\ell^-$ and $\ell^+$ are where $lf^{(\C)}_{B\ell}$ has fallen to $e^{-1/2}$
of its maximum value.  We remind the reader that
every sum over $\ell$ in this section is only over the values of
$\ell$ within band $B$.  

\subsection{Initial Binning}

One may wish to estimate fewer parameters than every multipole
moment right from the beginning.  In this case one would
parameterize the spectrum as 
\be
\label{eqn:Clparama}
{\cal C}_\ell = q_B {\cal C}_\ell^{\rm shape}\chi_B(\ell)
\ee
where $\chi_B(\ell)$ is one when $\ell$ is within the range 
of band $B$, and zero otherwise.  
 
To convert $q_B$ to a power estimate, ${\cal C}_B$, we need an average of the
shaped spectrum over band B.  A useful conversion factor
is given by Eq.~\ref{eqn:aveshape}.
Of course, in order to calculate $\C_B^{\rm shape}$ by Eq.~\ref{eqn:aveshape}
one needs to know the Fisher matrix at every $\ell$---which
is a calculation we're trying to avoid by using coarse binning.
Once again though, as long as the binning is not too coarse, the
details of the averaging are unimportant.  If the
binning is fine enough, then a simple average (uniform in $\ell$) 
will suffice---that is, take 
\be
\label{eqn:simpleaverage}
{\cal C}_B^{\rm shape} = {\sum_\ell 
{\cal C}_\ell^{\rm shape} \chi_B(\ell) \over \sum_\ell\chi_B(\ell)};
\ee
here, the denominator is simply the width of the bin.
This is what we have done in our applications (although see
\S~\ref{sec:RadComp} for how this can be improved by
use of analytic knowledge of the Fisher matrix).

As is usually the case with binning, we want to make the bins
as fine as necessary to capture all the information but
no finer since that means extra work.  A lower limit
to the bin sizes comes from the fact that fluctuation power from
${\cal C}_\ell$ will be indistinguishable from that
from ${\cal C}_{\ell'}$ if $|\ell - \ell^{'}| \lesssim 2 \pi/\theta$, where
$\theta$ is the linear extent of the observed region, as already mentioned.
We may wish to make our initial bins even coarser.  Some considerations
to keep in mind are that if one is trying to reduce sensitivity
to uncertainty in the power-law index then logarithmic spacing
produces equal shape sensitivity in each bin.  If the chief
shape uncertainty comes from features with a characteristic
wavelength, {\it e.g.}, Doppler peaks, then a linear spacing
produces equal shape sensitivity in each bin.  

\section{Application to Saskatoon}

We now apply our methods to the Saskatoon (SK) dataset\cite{nett95}.
The SK data are reported as complicated chopping patterns ({\it i.e.}, beam
patterns, $H$, above) in a circle of radius about $8^\circ$ around the
North Celestial Pole. The data were taken over 1993-1995 (although we
only use the 1994-1995 data) at an angular resolution of
$1.0$--$0.5^\circ$ FWHM at approximately 30~GHz and 40~GHz. More
details can be found in \cite{nett95}. The combination of the
beam size, chopping pattern, and sky coverage mean that SK is sensitive
to the power spectrum over the range $\ell=50$--$350$. 
The Saskatoon dataset is calibrated by observations of 
supernova remnant, Cassiopeia-A.  Leitch and collaborators \cite{Leitch}
have recently measured the flux and find that the remnant 
is 5\% brighter than the previous best determination.  We have
adjusted the Saskatoon data accordingly.  

\begin{figure}[bthp]
\plotonecita{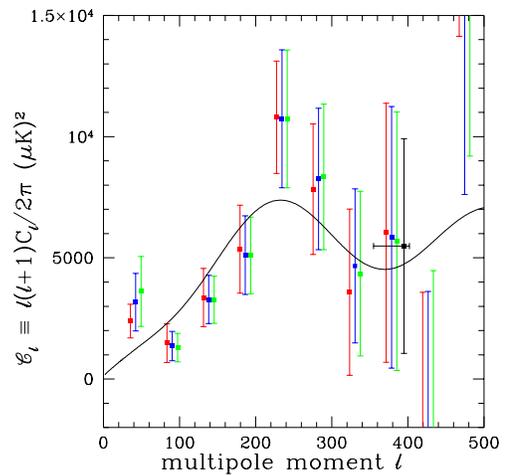}
\plotonecfpa{iterate_SK_quad0to3.eps}
\caption[SK quadratic iteration]{\baselineskip=10pt 
  Quadratic estimates of the power in 10 bins, derived from the SK data.
  The curve is the zeroth iteration, tilted CDM with $n = 1.45$ and
  $\sigma_8 = 2.16$.  The squares are from left to right, the results of
  the first to third iterations.  The data point with the horizontal
  error bar is a rebinning of the top three bins.}
\label{fig:SKiterate}
\end{figure}

In Fig.~\ref{fig:SKiterate} we show the results of our iterated
quadratic estimator on the SK data, in ten evenly spaced bins from
$\ell = 19$ to $\ell =499$.  Again, the
convergence proceeds quite rapidly, although not quite as
rapidly as for COBE/DMR. 
Evaluation of the Fisher matrix shows that there are approximately 
20\% anti-correlations between neighboring bins. We note in passing
that the falling power spectrum seen for $\ell\lesssim100$ has been noticed by
the experimenters themselves\cite{nettPriv}.

What we directly estimate is the adjustment factor $q_B$ of 
Eq.~\ref{eqn:Clparama}.  As mentioned above, in order to convert
this to a power spectrum amplitude we need some measure
of the average power in the bin.  Here we have used an average
uniform in $\C_\ell$ across the bin (Eq.~\ref{eqn:simpleaverage}).
For the first bin, the averaging should probably be weighted more
to the higher multipole moments than to the lower ones in the
bin because the sensitivity to the spectrum is increasing rapidly
with increasing $\ell$.
We will see this rapid rise in sensitivity to the power spectrum
in the next section where we plot the Fisher matrix for a finer binning.

There is very little information in the three highest $\ell$ bins.
Thus, for the final iteration we binned them together and plotted the
result as the point with the horizontal error bar.  Because of the
coarseness of the bins, the filter function for the rebinning is
coarse and therefore $\ell_{\rm eff}$, $\ell^{+}$ and $\ell^{-}$ are not
determined very well.  To get the filter function more finely, we
need to do a finer initial binning, which will be done in the
next section.  

\begin{figure}[bthp]
\plotonecita{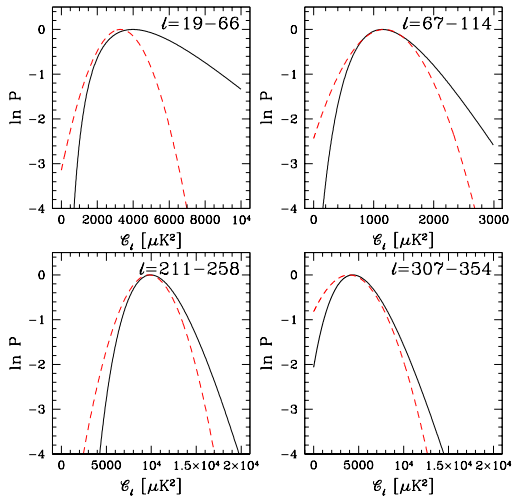}
\plotonecfpa{likefigs_SK.eps}
\caption[SK P(Cl)]{\baselineskip=10pt 
Probability Distributions for the power in bands, ${\cal C}_B$, as
labeled, for a prior uniform in $\C_B$.
The solid curve is the true likelihood from the 
direct evaluation; the dotted curve is the Gaussian approximation
from the third iteration of the quadratic procedure.}
\label{fig:SKP(Cl)}
\end{figure}

To investigate the probability
distributions beyond the mean and the variance, we used our direct
likelihood evaluation procedure, starting from the final quadratic
iteration.  The results are shown in Fig.~\ref{fig:SKP(Cl)}.  The
uncertainties in the first bin are strongly sample-variance dominated.
In the sample-variance limit the fractional variance, $(\delta
\C_\ell)^2/\C_\ell^2$, is inversely proportional to the number of
independent modes contributing to the estimate.  Since the first bin
is not well-determined we can therefore surmise that only a few modes
contribute to it.  With so few modes we cannot expect the
distribution to be Gaussian and thus the strong non-Gaussianity for
the first band, shown in Fig.~\ref{fig:SKP(Cl)}, is not surprising.

\section{Methods: Radical Compression}
\label{sec:RadComp}
As mentioned above, for Gaussian theories, $P(\Delta |\C_\ell)$
contains all the information that is in the map.  If the probability
distribution were Gaussian in $\C_\ell$, then all the information
in the probability distribution could be
compressed into a mean and a covariance matrix:
\be
P(\Delta|\C_\ell) \rightarrow \hat \C_\ell$, $\langle \delta \C_\ell 
\delta \C_{\ell'}\rangle.
\ee
By the definition of a Gaussian probability distribution, this
compression involves no loss of information.  The ``lossless'' nature
of this compression was pointed out by Tegmark \cite{tegmarkoptimal}
although here we emphasize that it is only true in the Gaussian limit.
We refer to compression to the power spectrum as ``radical
compression'' because the data reduction is impressive: the
information in a map with $N$ pixels and an $N \times N$ noise
covariance matrix is now held in less than $\sqrt{N}$ power estimates
and their $\sqrt{N} \times \sqrt{N}$ covariance matrix.

With compression to $\sqrt{N}$ numbers and a covariance matrix, analysis
of constraints on cosmological parameters becomes quite rapid.
One simply forms the $\chi^2$:
\be
\label{eqn:chisqanalysis}
\chi^2(\{a\}) = \sum_{\ell\ell'}\left(\C_\ell(\{a\}) - \hat\C_\ell\right)
M^{-1}_{\ell \ell'} \left(\C_{\ell'}(\{a\}) - \hat\C_{\ell'}\right)
\ee
and simply evaluates it to find the minimum and also the one sigma and
possibly two sigma confidence regions of the parameter space.  Here,
$\C_\ell(\{a\})$ is the calculated spectrum for the paramters $a_p$ and
$M_{\ell\ell'}\equiv\langle \delta \C_\ell \delta \C_{\ell'}\rangle$ is
some appropriately determined correlation matrix, {\it e.g.}, the
inverse of the Fisher matrix or the exact curvature matrix for the
quadratic method, or a likelihood ratio or Bayesian determination for
the direct evaluation of the likelihood.

Unfortunately, the probability distribution is non-Gaussian, as we have
seen.  One might think that this only causes minor inaccuracies to the
method of Eq.~\ref{eqn:chisqanalysis}.  In fact, the problems are of a
systematic nature and can be quite important.  To see this we need only
examine the case of COBE/DMR.  Say we wanted to use our power spectrum
estimates to measure the best fit amplitude of standard CDM, expressed
as a prediction for $\sigma_8$, by using Eq.~\ref{eqn:chisqanalysis}.
Using our estimates of $\C_\ell$ from the final iteration of either the
direct or quadratic estimation procedures together with the Fisher
matrix from the final iteration, we find $\sigma_8 = 1.1$ instead of the
correct value of $\sigma_8 = 1.2$.  This example does not mean that
non-Gaussianity has made radical compression useless, but rather that we
must proceed with some care.

The decrease in power is a systematic effect due to the skewness of the
probability distributions which allow more positive and less
negative fluctuations relative to a Gaussian distribution with the
same variance.  Another way of thinking about it is that those
amplitudes that fluctuate downwards have their variance reduced and thus
their weight increased while those that fluctuate upward have their
variance increased and therefore their weight decreased.  Contrast this
to a Gaussian probability distribution for which the curvature is
independent of location. Thus one can see that the non-Gaussianity of
the probability distribution can be very important and some care must be
used in attempting this radical compression.

One solution to the problem may be to find a function of $\C_\ell$ whose
distribution is more Gaussian than that of $\C_\ell$ itself.  Motivation
for one particular form comes from considering the sources of the
variance.  There is a sample-variance contribution which is proportional
to the power and a noise contribution which is independent of the power,
thus $\delta \C_\ell \propto \C_\ell + x_\ell$ for some appropriate $x_\ell$
related to the experimental noise.  According to this
proportionality, the probability distribution for $\ln\left(\C_\ell +
  x_\ell\right)$ might be well-approximated by a Gaussian since its variance
is independent of $\C_\ell$.  This procedure is under investigation
\cite{ourotherpaper}.

It is the iterated Fisher
matrix that overweights (underweights) the points that 
fluctuated downward (upward); to prevent these fluctuations
from affecting the Fisher matrix, one can iterate on
the parameters of a smooth function, instead of the amplitudes
in fine bins, and then use the resulting Fisher matrix for the
covariance matrix associated with the power estimates in bins.
This is the method of solution we have adopted here.  

We emphasize that the problems we are discussing are not
peculiar to the use of the quadratic estimator, but are associated
with the attempt to compress the probability distribution of
$\C_\ell$ into a mean and covariance matrix.  Because
of non-Gaussianity, this procedure is necessarily approximate.  
The above being said, we will now assume Gaussianity, but always
use the Fisher matrix derived from a smooth theory curve, and
not one derived from a bin by bin iteration.

A more benign problem than non-Gaussianity is the existence of
correlated uncertainties.  Although not a problem for the
$\chi^2$ of Eq.~\ref{eqn:chisqanalysis}, the correlations do
complicate direct visual interpretation.  We may remove these
correlations by a linear transformation on the parameter space,
$q \rightarrow \tilde q = Z q$, where $Z$ diagonalizes the
parameter covariance matrix, $ZF^{-1}Z^T = diag$ (or,
equivalently, $Z^{-1}$ diagonalizes the Fisher matrix,
$F^{(q)}$).

While having the advantage of uncorrelated uncertainties, the
interpretation of these new parameters themselves has been complicated
by the transformation.  Fortunately, there are transformations with
some very useful properties.  Many transformations can lead to
independent modes.  Hamilton\cite{Hamilton} made a lengthy study of
the different possible diagonalizing transformations and found one
that has a small $\ell$-space width and is positive-definite.  His
transformation translates to $Z = L^T$ where $L$ is the Cholesky
factorization of $F$; $F = LL^T$.  Another useful transformation is
given by setting $Z = F^{1/2}$, the Hermitian square root of
$F$\cite{TegHam}. Parameter eigenmodes\cite{bet97}, involving rotation
only to the orthogonal combinations of bandpowers, rather than scale
transformations as well, are of great interest and emphasize another
point: when linear combinations of modes are being taken, we have
freedom in exactly what the scaling will be. It is obvious though that
if we are representing a bandpower at a representative $\ell$, we want
to ensure that the normalization makes sense since the goal is direct
visual comparison with theoretical ${\cal C}_\ell$ curves. 

In the Gaussian approximation, these linear combinations are
independent.  Thus we can now estimate each ${\tilde q}_j$
independently of the ${\tilde q}_i$ for $i \ne j$.  The
$\ell_{max}$-dimensional space has been reduced to $\ell_{max}$
one-dimensional spaces.  In fact, for each ${\tilde q}_j$, there is no
need to keep the Gaussian approximation; one can calculate its
complete distribution function.  If the Fisher matrix is a good
approximation to the curvature matrix, then, at least near the peak,
the total likelihood function can be approximately decomposed into a
product of these one-dimensional likelihood functions: ${\cal
L}(\{{\tilde q}_i\})\approx\prod_i{\cal L}({\tilde q}_i)$.  Since
$\tilde q_\lambda = q_\ell Z_{\ell\lambda} = {\cal C}_\ell/{\cal
C}_\ell^{\rm shape} Z_{\ell \lambda}$, the filter function is
$f^{(\C)}_{\lambda \ell} = Z_{\ell \lambda}/{\cal C}_\ell^{\rm
shape}$.  Thus \be {\cal C}_{\lambda}={\tilde q_\lambda \over
\sum_\ell f^{(\C)}_{\lambda \ell}} = {\sum_\ell{\cal C}_\ell
f^{(\C)}_{\lambda \ell}\over \sum_\ell f^{(\C)}_{\lambda \ell}}.  \ee

If we wish to rebin these uncorrelated estimates, we can
do so in a minimum variance manner by performing the following
sums:
\be
\label{eqn:calCbeta}
{\cal C}_\beta = {\sum_{\lambda \in \beta} \tilde q_\lambda N_\lambda
 \over \sum_\ell f^{(\C)}_{\beta \ell}}
\ee
where
\be
f^{(\C)}_{\beta \ell} = \sum_{\lambda \in \beta} f^{(\C)}_{\lambda \ell} N_\lambda
\ee
and $N_\lambda \equiv \sum_\ell Z_{\ell \lambda}$.  These
equations are derived in Appendix B.  

\begin{figure}[bthp]
\plotonecita{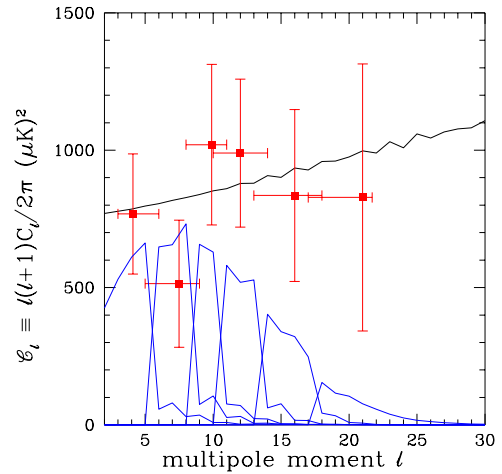}
\plotonecfpa{cobe_rebin_quad4_fish0.eps}
\caption[DMR with ortho-rebinning]{\baselineskip=10pt 
DMR with ortho-rebinning. }
\label{fig:DMRfinal}
\end{figure}

For COBE/DMR we used Cholesky decomposition to get filter functions,
$f^{(\C)}_{\lambda \ell} = L_{\ell\lambda}/\C^{\rm shape}_\ell$, and then
binned together combinations 1-3,4-5,6-7,8-9,10-12,13-16 and 17-27.
Each of the combinations' filter functions is shown in 
Fig.~\ref{fig:DMRfinal}, together with the power estimates.  To
avoid the systematic underestimate of power discussed above
we used the $\C_\ell$s from our final iteration, but the
Fisher matrix from the zeroth iteration.  This does not imply
that the errors are completely unaffected by the data.  We
are using standard CDM as our zeroth iteration precisely 
because it is a good fit to the data.

\begin{figure}[bthp]
\plotonecita{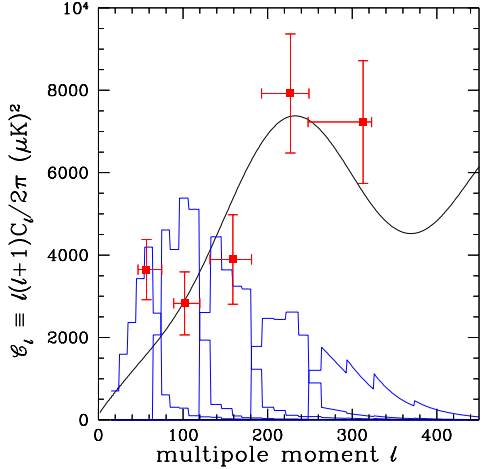}
\plotonecfpa{skfinal.eps}
\caption[SK with ortho-rebinning]{\baselineskip=10pt 
SK with ortho-rebinning.}
\label{fig:SKfinal}
\end{figure}

In order to make accurate filter functions for SK we divided
it up into 26 bins.  Starting from the results of our third 
iteration on the 10 bands
of the previous section we estimated the power in these 26 bands
with a single iteration.
The fractional uncertainty in most of these bands was greater
than unity.  In order to make the rebinned 
statistically orthogonal linear combinations plotted in Fig.~\ref{fig:SKfinal} 
we used the $n=1.45$, $\sigma_8 = 2.16$
tilted standard CDM Fisher matrix.  These are the parameter values
for tilted CDM that maximize the likelihood function.

At high $\ell$, some of our 26 bands still have significant
width; their ``sub-band structure'' may be important.  To estimate
the structure of the filter functions {\it within}\/ each band,
we employ an analytic approximation to the Fisher matrix.   
For a map of the sky with uniform weight per solid angle, $w$,
covering a fraction of the sky, $f_{\rm sky}$, 
we know the Fisher matrix is such that
\cite{knox95,forecast,bet97}:
\be
\label{eqn:mapfishmat}
\sum_{\ell'}F^{(\C)}_{\ell\ell'} \simeq {(2\ell+1)f_{\rm sky}\over 2}
\left[{\cal C}_\ell+{\ell(\ell+1)\over 2\pi w {\cal B}^{2}(\ell)}
\right]^{-2}, 
\ee 
where, for a Gaussian beam, ${\cal B}(\ell) =
e^{-\ell^2\sigma_b^2/2}$.  

An approximation appropriate for difference experiments rather than
maps is to replace $w{\cal B}^2(\ell)$ with the noise-weighted window
function $w{\overline W}^{(N)}_\ell \equiv {\rm Tr}(C_N^{-1}{\bf
W}_\ell)/N$, where 
${\bf W}_\ell$ is the window function
matrix of Eq.~\ref{eqn:wl}.  In this uniform case, we also have
\begin{eqnarray}
\label{eqn:fishqhomnoise}
&& f^{(q)}_{B\ell} \simeq {(2\ell+1)f_{\rm sky}\over 2}
\Big[{\eps_{T\ell}\over (1+\eps_{T\ell})}\Big]^2\, , \nonumber \\ 
&&
\eps_{T\ell} \equiv w{\overline W}^{(N)}_\ell {\cal C}_\ell {2\pi
\over \ell(\ell+1)} \, .  
\end{eqnarray}
The quantity $\eps_{T\ell}$ is a measure of the mean square of the
signal-to-noise ratio in modes $\ell$. The $\eps_{T\ell}/
(1+\eps_{T\ell})$ factor which appears in the square is the Wiener
filter ({\it i.e.}, the optimal signal-to-noise filter). In the
$\eps_{T\ell} \gg 1$ limit of signal-dominance,
$f^{(q)}_{B\ell}\to(\ell+1/2)f_{\rm sky}$, half the number of $\ell$
modes available. This is of course a general result for the
signal-dominated regime, requiring no assumption of homogeneous noise.
It is often a reasonable approximation to use the usual filter
function $W_\ell \equiv {\rm Tr}({\bf W}_\ell)/N$ in place of the
noise-weighted one. Eq.~\ref{eqn:fishqhomnoise} is applied to
realizations of power spectra for future balloon and satellite
experiments in \S~\ref{sec:forecast}. 

\begin{figure}[bthp]
\plotonecita{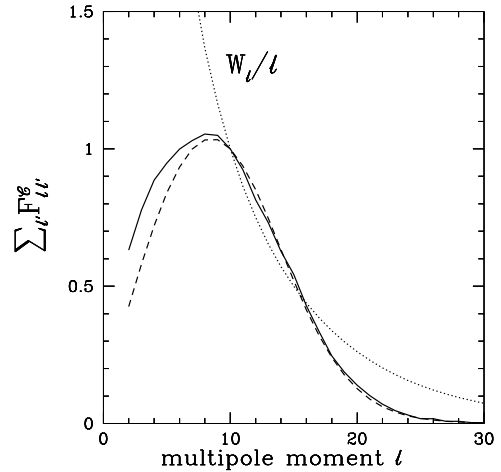}
\plotonecfpa{dmrfilters_1band.eps}
\caption[DMR Fisher matrix]{\baselineskip=10pt 
The Fisher matrix sum, $\sum_{\ell'}F^{(\C)}_{\ell\ell'}$ for the
zeroth iteration of DMR for exact (solid) and analytic (dashed).
For comparison, $W_\ell/\ell$ (dotted) is also shown.}
\label{fig:DMRfilt}
\end{figure}

The weight map for COBE/DMR varies gently with spatial scale outside
of the galactic cut, so we expect the analytic approximation
eq.~\ref{eqn:mapfishmat} to be reasonably good for it and we see in
Fig.~\ref{fig:DMRfilt} that this is so.  The two curves with a peak at
$\ell = 10$ are sums over the exact and analytic Fisher matrices for
standard CDM.  For the analytic form we took $f_{\rm sky} = 0.65$,
$w^{-1} = 9.5 \times 10^{-13}$ (equivalent to an rms noise of $22 \
\muK$ on $7^\circ$ pixels) and the appropriate beam shape.

\begin{figure}[bthp]
\plotonecita{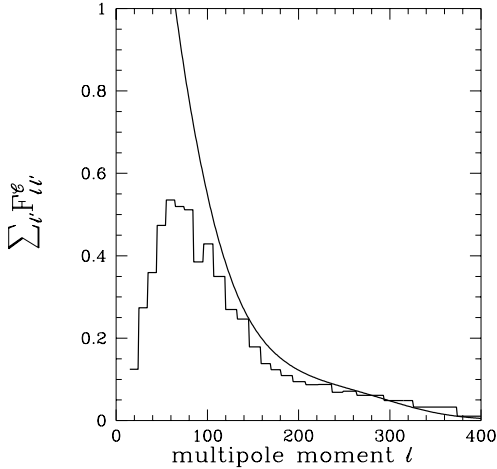}
\plotonecfpa{SK26filt_exact_ana.eps}
\caption[SK26 Fisher matrix]{\baselineskip=10pt 
The Fisher matrix sum, $\sum_{\ell'}F^q_{\ell\ell'}$ for the
zeroth iteration on 26 band SK for exact and analytic.}
\label{fig:SK26filt}
\end{figure}

For the SK data, the comparison of the 26 band exact Fisher matrix and
the analytic Fisher matrix approximation shows some interesting
differences (Fig.~\ref{fig:SK26filt}).  The analytic curve is for
$f_{\rm sky} = 0.005$, $w^{-1} = 3.3 \times 10^{-14}$ and $\theta_{\rm
  fwhm} = 0.5^\circ$.  The deficit at smaller $\ell$ is presumably due
to the differencing schemes that were necessary to filter out
atmospheric contamination. These are partly encoded in the
noise-weighted ${\overline W}^{(N)}_\ell$, but for the plot only the
beam, ${\cal B}^2(\ell)$, was used, as in Eq.~\ref{eqn:mapfishmat}.
This deficit bears on the question of what quality of map it is possible
to create from the SK dataset; the loss of low $\ell$ information
implies that there will be long-distance noise correlations in any map
made from the data \cite{SKmap}.

\begin{figure}[bthp]
\plotonecita{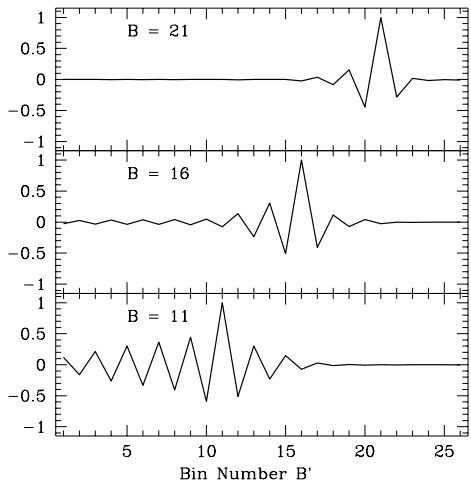}
\plotonecfpa{skcov_11_16_21.eps}
\caption[SK covariance matrix]{\baselineskip=10pt 
Slices of the SK 26 band covariance matrix at
bands 11, 16 and 21, normalized to unity along the diagonal.}
\label{fig:SK26cov}
\end{figure}

In Fig.~\ref{fig:SK26cov} we show some rows of the normalized parameter
covariance matrix $M_{BB'}/\sqrt{M_{BB}M_{B'B'}}$, where
$M_{BB'}=F^{-1}_{BB'}$.  The correlations for bin 11 ($\ell=120$--$132$)
extend well beyond the $\delta \ell \simeq \pi/\theta$ expected for a
map---again, this is presumably due to the differencing schemes.

Returning to Fig.~\ref{fig:SK26filt}, we see that the agreement at least at higher $\ell$, 
is good.  We consider it to be good enough
to encourage the use of the analytic form for doing some
sub-band shaping of the filter functions.  To be more precise
about the procedure, within each band, $B$, we give $f^{(\C)}_{\beta
\ell}$ the shape $\sum_{\ell'}f^{(\C)}_{\ell \ell'}\C_\ell$
with the amplitude chosen so that $f^{(\C)}_{\beta B} =
\sum_\ell f^{(\C)}_{\beta \ell}$.  We have applied this shaping to the
five highest $\ell$ bins in Fig.~\ref{fig:SKfinal}.  

One might wonder why the analytic curve in Fig.~\ref{fig:SK26filt} has
no peak, corresponding to where sample variance and noise are equal
contributors to the uncertainty in $\C_\ell$.  The absence of the peak
is due to the rise in $\C_\ell$ from $\ell = 20$ to $\ell =200$.  If we
plotted $\C_\ell^2\sum_{\ell'} F^{(\C)}_{\ell \ell'}$, which is related
to the {\it fractional}\/ uncertainty in $\C_\ell$, then there would be a
peak near $\ell = 80$.

While the independence of the power estimates (in the Gaussian
approximation) simplifies Eq.~\ref{eqn:chisqanalysis} some,
the existence of the filter functions complicates it:
\be
\label{eqn:chisqanalysis2}
\chi^2(\{a\}) \equiv \sum_B {1\over(\delta \C_B)^2}
\left[{\C}_B(\{a\}) - \hat{\C}_B\right]^2\,  ,   
\ee 
where ${\C}_B(\{a\})$ is calculated using the filters in
Eq.~\ref{eqn:aveshape}. 
We have cast this equation in an intuitive form involving the
deviation of a measured bandpower ${\cal C}_B$ from the predicted
spectrum. This is exactly the $\chi^2$ appropriate to $q$-space, which
emphasizes relative deviations of both the data and the theoretical
predictions from the fiducial spectrum used to calculate the quadratic
estimator; {\it i.e.}, the details of how one goes from the estimates
of $q_B$  to the appropriate bandpower drop out of the $\chi^2$.

One might argue that the complication of the covariance matrix has
been traded for the complication of the filter functions and there has
been no net improvement.  However, we think that, when binning has
been done, the use of the orthogonal linear combinations improves, or
at least simplifies, the process of radical compression.  Once binning
has occurred, one wants to know what the filter looks like across the
bin.  Thus binning implies the use of filters and once filters are
being used, the orthogonal linear combination approach of providing
uncorrelated data and filters is simpler than providing correlated
data with filters {\em and}\/ a covariance matrix.

Experiments typically report broad-band power spectrum estimates,
together with the trace of their window function, $W_\ell$, which can
be used to make a filter function, $f_\ell = W_\ell/\ell$.  These
power spectrum estimates have indeed been used to constrain
cosmological parameters, {\it e.g.}\/ \cite{Lineweaver}.  Using
$f_\ell = W_\ell/\ell$ as the filter function is in general not the
optimal procedure.  Only if $C_{Tii'}$ is a multiple of the identity
matrix is $W_\ell/\ell$ the minimum-variance filter.  In general, the
Fisher matrix-derived filters should be used.  And they can be quite
different; in Fig.~\ref{fig:DMRfilt} one can see the tremendous
difference between $W_\ell/\ell$ and the minimum variance filter,
$f_\ell$.  In the noise-dominated regime (high $\ell$ for DMR),
$W_\ell/\ell \propto \ell^{-1}B^2(\ell)$ whereas $f_\ell \propto
\ell^{-3} B^4(\ell)$.

In our power spectrum plots we have not included calibration
uncertainty which is $\sim 6\%$ for SK and negligible for COBE/DMR.  The
calibration uncertainty is completely correlated across the bands.  It
can be taken into account as a nuisance parameter to be added to the
$\chi^2$ expression above \cite{Lineweaver}.  Other methods for taking
it into account are discussed in \cite{ourotherpaper}.

\section{Forecasting Power Spectra for Future Experiments}\label{sec:forecast}

\begin{figure}[bthp]
\noindent\hskip-0.6cm{
{\vbox{\epsfxsize=10cm
\epsfboxcita{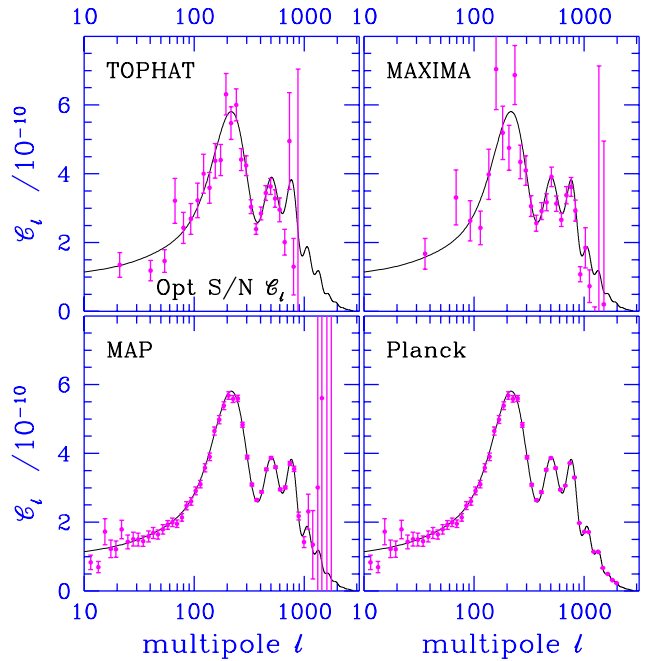}
\epsfboxcfpa{bjk97_figoptStoNCL.eps}
}}}
\vskip-0.2cm 
\caption[Signal-to-noise power spectrum forecasts]
{Comparison of forecasts for the two balloon experiments, TOPHAT
and MAXIMA, with the satellite experiments MAP and Planck. Bands are
required to have a signal-to-noise of at least 25 and a minimum
spacing in $\ell$ defined by the logarithmic spacing $\Delta \ln \ell
=0.1$. With this signal-to-noise binning, the growth in the
number of bands shows the increasing precision and sky coverage
of the experiments. The error bars are those appropriate to the 
quadratic estimator after convergence. }
\label{fig:CLB_LDBcfSAT}
\end{figure}

\begin{figure}[bthp]
\noindent\hskip-0.6cm{
{\vbox{\epsfxsize=10cm
\epsfboxcita{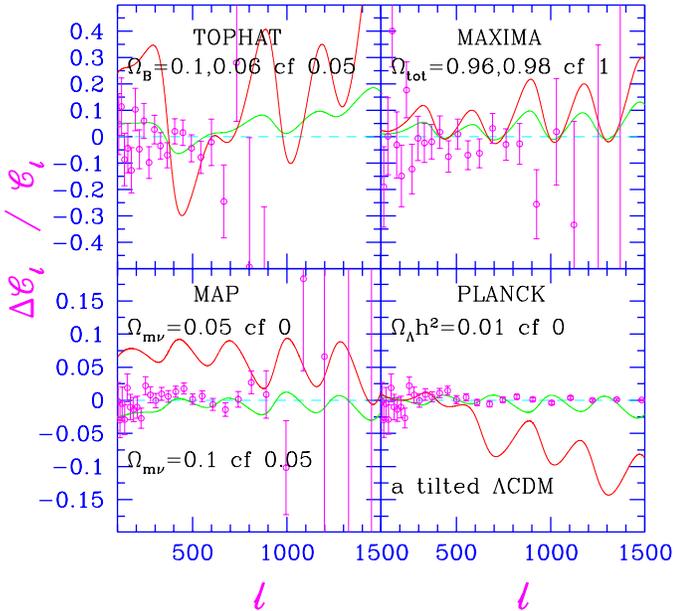}
\epsfboxcfpa{bjk97_figoptStoNq.eps}
}}}
\caption[Signal-to-noise power difference forecasts] {The forecasted
  data with error bars are shown in $(q=\Delta\C_\ell/\C_\ell)$-space,
  in which the relevant comparison with the data is the fractional
  difference between the ${\cal C}_\ell$ we are testing and ${\cal
    C}_\ell^{shape}$. A few differences are shown for each case by solid
  lines. They are deviations in single parameters, as marked, from the
  shape theory ($q=0$), in this case a standard COBE-normalized CDM
  model with $\Omega_B=0.05$.  The theoretical curves can have their
  amplitudes adjusted up or down to best fit the simulated data. }
\label{fig:sigth2LB_LDBcfSAT}
\end{figure}

In this section, we exercise our methods on an instructive simple
case, homogeneous noise over regular patches covering a fraction
$f_{\rm sky}$ of the sky. We apply the relations to simulating
realizations of power spectra and their error bars for two planned
balloon experiments, MAXIMA and TOPHAT, and two
satellite experiments, MAP and PLANCK. The results are shown in the
familiar ${\cal C}_\ell$ space in Fig.~\ref{fig:CLB_LDBcfSAT} and in
$\Delta {\cal C}_\ell /{\cal C}_\ell$ space in
Fig.~\ref{fig:sigth2LB_LDBcfSAT}. In this $q$-space, which we believe
will become more and more utilized as the CMB data starts to converge
on a specific shape, we compare the (converged) quadratic power
estimator values and their error bars with the fractional deviation,
$\Delta {\cal C}_\ell /{\cal C}_\ell$, of a ${\cal C}_\ell$ whose
parameters we are testing from a fiducial shape. Here the shape that
entered into the power spectrum analysis was a standard
COBE-normalized cold dark matter model ${\cal C}_\ell$, and the model
used to construct the power spectrum realization was also this SCDM
one. 

For $f_{\rm sky} = 1$, power spectrum analysis simplifies considerably
if the weight matrix $C_N^{-1}$ is diagonal in the spherical harmonics
basis, since then the $C_T$ and $C_{T,p}$ matrices are. Both the Fisher and
curvature matrices are also diagonal as long as the bands $B$ do not
overlap in $\ell$-space. For $f_{\rm sky}< 1$, another simple limiting
case involves rectangular regions of size $N_x\pomega_{pix}\times
N_y\pomega_{pix}$, consisting of square pixels of size
$\pomega_{pix}\times \pomega_{pix}$. The S/N eigenmodes are then
discrete Fourier components, labelled by a wavevector ${\bf Q}$,
which, to a high degree of accuracy, diagonalize $C_T$ and $C_{T,p}$,
and, by assumption, $C_N^{-1}$. The number of modes of a given $\vert
{\bf Q} \vert $ available in a $d\vert {\bf Q} \vert =1$ band is
$(N_xN_y\pomega_{pix}^2 /(2\pi))\vert {\bf Q} \vert $; {\it i.e.}, $f_{\rm
sky}2\vert {\bf Q} \vert $. Using $\vert {\bf Q} \vert \approx \ell
+\half$, which follows from relating an expansion in these modes to an
expansion in spherical harmonics at high $\ell$ \cite{bh95}, the
number of modes is $(2\ell +1) f_{\rm sky}$, as in the all-sky case.

The Fisher matrix and the quadratic $q$-estimator are given by 
\begin{eqnarray}
&& \delta Q_B = v_B /(2F)_{BB} \, , \quad F_{BB} = \sum_{\ell \in B}  f^{(q)}_{B \ell} \, , 
\label{eqn:fishquadhomnoise} \\
&& v_B =  \sum_{\ell \in B} 2 f^{(q)}_{B\ell} {\{(1+\eps^{{\rm true}}_{T\ell})
\rho_\ell^2 -(1+ \eps^{(*)}_{T\ell}) \} \over \eps^{(*)}_{T\ell}} \, , \nonumber \\
&& 2 f^{(q)}_{B\ell} \equiv g_\ell \Big[\eps^{(*)}_{T\ell}
/(1+\eps^{(*)}_{T\ell})\Big]^2 \, ,  \quad g_\ell \equiv (2\ell +1) f_{\rm sky} \, . \nonumber
\end{eqnarray}
The signal-to-noise factor $\eps_{T\ell}$ is related to the average
weight $w$, the noise-weighted filter function ${\overline W}_\ell$,
and ${\cal C}_\ell$ by Eq.~\ref{eqn:fishqhomnoise}; and the expression
for $f^{(q)}_{B\ell}$ is a repeat of Eq.~\ref{eqn:fishqhomnoise}. It
is also straightforward to modify $\eps_{T\ell}$ to take into account
the noise in multifrequency experiments, including the expected beam
size variation with frequency channel\cite{bet97}.

The combination $(1+\eps^{{\rm true}}_{T\ell})\rho_\ell^2$ is the
average power in the modes with given $\ell$, where $\eps^{{\rm
true}}_{T\ell}$ is the true value of the power spectrum, and
\begin{eqnarray}
&& \rho_\ell^2 = g_\ell^{-1} \sum_{\mu \in \{ \ell-{\rm modes} \}}
{\rm GRD}_{\ell\mu}^2 \, , \nonumber 
\end{eqnarray}
where ${\rm GRD}_{\ell\mu}$ is a Gaussian random deviate for the mode of
given $\ell$ labelled by a degeneracy variable $\mu$ (the azimuthal
quantum number, $m$, in the spherical harmonic case, a discrete angle index in
the rectangular patch case); individual realizations of this variable
are due to sample and/or cosmic variance. Therefore, $g_\ell \rho_\ell^2$ is
distributed like $\chi^2$ with $g_\ell$ degrees of freedom, {\it i.e.},
with a cumulative probability given by an incomplete Gamma function with
arguments $g_\ell/2$ and $\rho_\ell^2/2$. Numerical realizations can be
done very quickly.

The factor $\eps^{(*)}_{T\ell}$ denotes an approximate value for the
signal-to-noise power spectrum. As we have discussed, within the band
we adopt an assumed shape but allow the amplitude $Q_B$ to vary. 
In the iterative scheme, $\eps_{T\ell}^{(*)}=\eps^{(n)}_{T\ell}$ would be the
value on iteration $n$, and $\eps_{T\ell}^{(n+1)}=(1+\delta Q_B)
\eps_{T\ell}^{(n)}$ would be the value to be inserted for the next
iteration. 

Note that $Q_B$ is the weighted average of the quadratics in sub-bands
of width unity, $Q_\ell$, with weight $ f^{(q)}_{B \ell}$. Therefore,
the classic optimal signal-to-noise filter, the Wiener filter,
$\eps_{T\ell}/(1+\eps_{T\ell})$ in this case, enters in a fundamental
way into the power spectrum estimation procedure.

In the signal-dominated region, $\eps_{T\ell} \gg 1$, the weighting is
just by number of modes, $2f^{(q)}_{B\ell} \rightarrow g_\ell$. Thus
$F$ does not change, and $\delta Q_B$ converges after one
iteration. The ${\cal C}_\ell$ error bars change because the
$\avrg{{\cal C}_\ell}_B$ is multiplied by the converged $(1+\delta
Q_B)$. In the fine-grained case, where $B$ encompasses just one
$\ell$, the $ f^{(q)}_{B\ell}$ weights in the
$v_B$ numerator and the Fisher denominator cancel, leaving
$1+\eps_{T\ell}^{(n)}$ = $(1+\eps^{{\rm true}}_{T\ell})\rho_\ell^2$
for $n \ge 1$ even in the noise-domimated regime. 

We adopt improved specifications especially in beam size for MAP
\cite{map96ref} and PLANCK \cite{cobrassamba96ref} over the original
proposal values; these are likely to evolve for Planck.  Of the 5 HEMT
channels for MAP, we assume the 3 highest frequency channels, at 40,
60 and 90 GHz, will be dominated by the primary cosmological signal
(with 30 and 22 GHz channels partly contaminated by bremsstrahlung and
synchrotron emission). MAP also assumes 2 years of observing. For
Planck, 14 months of observing and current (proposal-modified) values
are used. The HEMT-based LFI specifications are significantly
improved; the 100, 65, 44 GHz channels, but not the 30 GHz channel,
were used. For the bolometer-based HFI, 100, 150, 220 and 350 GHz were
used. Dust-contamination will certainly affect the 550 and 850 GHz
channels. For both, it was assumed that 65\% of the sky would be
useful. MAP has $w^{-1}=0.8\times
10^{-15}$ and PLANCK has $w^{-1}=3.3\times 10^{-18}$.

The balloon forecasts used conservative numbers for
the bolometer-based TOPHAT \cite{TopHat97} and MAXIMA \cite{MAXIMA97}
experiments that take account of excess noise associated with
foreground removal. It was assumed that 65\% of the region covered by
TOPHAT would be useable for CMB analysis ($f_{\rm sky}=0.028$). The beam
is $20^\prime$ and $w^{-1}=1.5\times 10^{-15}$ was chosen. (These noise
values are for roughly a 10 day mission.) MAXIMA has a $12^\prime$ beam,
and $f_{\rm sky}=0.01$, $w^{-1}=0.9\times 10^{-15}$ were chosen.

Other long duration balloon (LDB) bolometer experiments such as
Boomerang \cite{Boomerang97} should be able to do as well. HEMT-based
LDB experiments, such as BEAST \cite{ACE97} using 40 GHz HEMTs, might
also achieve similar accuracy. A sharp lower $\ell$-cut was included
to treat the limited sky coverage for TOPHAT ($\ell_{cut}=12$) and
MAXIMA ($\ell_{cut}=20$); we allowed one mode per $\ell$ above this
until $(2\ell +1) f_{\rm sky}$ exceeded unity, at which point the
number of modes was given by the integer part of $(2\ell +1) f_{\rm
sky}$. An uncertain part of this approach is the treatment of modes of
order the size of the patch.

In Fig.~\ref{fig:CLB_LDBcfSAT}, we have tested various prescriptions
for placing the power and the $\ell$ value. In \S~\ref{subsec:rebin},
we recommended using $f^{({\cal C})}_{B \ell} = 2f^{(q)}_{B\ell}/{\cal
C}_\ell $, but other schemes can also be defended; {\it e.g.}, 
weighting by the power in the modes, so the numerator averages
$[\ell(\ell+1)]^{-1} {\cal C}_\ell $ {\it wrt} $f^{(q)}_{B\ell}$ and
the denominator averages $[\ell(\ell+1)]^{-1}$. For a steeply falling
spectrum, the former places the error bar at high $\ell$, with power
weighting it is placed at slightly lower $\ell$.  In all cases,
$f^{(q)}_{B\ell}$ is essential to include, but, apart from this, the
main lesson we have learned is that otherwise the prescription does
not matter very much.

The decision on the number and placement of bands has also been
explored. We prefer using a combination of conditions to determine the
spacing: when the S/N estimate $v_B/(2\sqrt{F_{BB}})$ exceeds some
threshold, or if $\Delta \ln \ell$ across the band reaches some
precribed value, then a new band is made. If we only used logarithmic
spacing, then there would be too many bands at lower $\ell$ with
poorly determined bandpowers for TOPHAT and MAXIMA. For the figures,
we chose a S/N minimum of 25, translating to a 20\% fractional error
on ${\cal C}_\ell^{1/2}$; we also chose $\Delta \ln \ell
=0.1$. Clearly, because of the all-sky nature of MAP and Planck, the
bands are mostly determined by the logarithmic criterion. This is only
true at the higher $\ell$ (but before the beam kicks in) for the balloon
experiments.

One of the nice features of the homogeneous sky simplicity is that we
can easily test what different prescriptions and weightings will
do. For example, we have explored other ways of finding the maximum
and estimating the errors. The nonlinear maximum likelihood estimator
uses the curvature matrix:
\begin{eqnarray} 
&& \delta Q_B ({\rm max}{\cal L}) =v_B /(2{\cal F})_{BB} \, , \label{eqn:NLmaxLhomnoise}\\
&&{\cal F}_{BB} =F_{BB} +  \sum_\ell 2 f^{(q)}_{B \ell}
\ {(1+\eps^{{\rm true}}_{T\ell})
\rho_\ell^2 -(1+ \eps^{(*)}_{T\ell}) \over (1+\eps^{(*)}_{T\ell})} 
\, .   \nonumber 
\end{eqnarray}
In the fine-grained case, the amplitude adjusts until
$1+\eps_{T\ell}^{(n)}$ = $(1+\eps^{{\rm true}}_{T\ell})\rho_\ell^2$,
so ${\cal F}_{BB} \rightarrow F_{BB}$ and the two power spectrum
estimates and their error bars are the same. This form, though, takes
longer to converge than the quadratic, and when the deviations are too
large the iteration may not converge. (This is typical for the
Newton-Raphson method.) A comparison of ${\cal F}_{BB}$ in
Eq.~\ref{eqn:NLmaxLhomnoise} and $v_B$ in
Eq.~\ref{eqn:fishquadhomnoise} shows that, for wider bands, we can
expect plus and minus fluctuations over the band which give $v_B =0 $,
but, because of the different weighting, will not give ${\cal F}_{BB}
=F_{BB}$. 

For the quadratic operator, another measure of the error bars is the
variance of the $Q_B$, and this can partly take the non-Gaussian
spread of the probability function for the quadratic into account. For
the case considered here, this variance is diagonal in $B$. When the ensemble
average is taken, the result is
\begin{eqnarray}
&& \avrg{ \Delta Q_B \Delta Q_B 
} \nonumber \\
&& = 
F_{BB}^{-1} \ {\sum_{\ell \in B} f^{(q)}_{B\ell}
{(1+\eps^{{\rm true}}_{T\ell})^2/ (1+\eps^{(*)}_{T\ell})^2} \over \sum_{\ell\in B}
f^{(q)}_{B\ell} } \,
\nonumber \\
&&+ \Big[{\sum_\ell f^{(q)}_{B\ell}( \eps^{{\rm true}}_{T\ell}/\eps^{(*)}_{T\ell} -1) \over 
\sum_\ell  f^{(q)}_{B\ell}}\Big]^2
\, . 
\end{eqnarray}
Thus in the limit that $\eps^{(*)}_{T\ell}$ approaches $\eps^{{\rm
true}}_{T\ell}$, it reduces to $F_{BB}^{-1}$, the Fisher error we
quote. However, the $\rho_\ell^2$ corrections inherent in any
realization preclude convergence to $F_{BB}^{-1}$, in such a way as to
increase the error bars for low power modes and lowering them for high
power modes over what $F_{BB}^{-1}$ gives. 

\section{Discussion}

\subsection{Computer Resource Demand}

Evaluating the likelihood function is an $O(N^3)$ operation.  Although
both matrix inversion and determinant calculation are $O(N^3)$
operations, it is only the determinant evaluation that prevents the
likelihood analysis from being $O(N^2)$.  
That is because only $C^{-1}\Delta$ is needed in the $\chi^2$
evaluation, not the full inverse, and this can be potentially be
calculated via $O(N^2)$ iterative techniques.  Today, a {\em single
  evaluation}\/ of the likelihood function (and it must be evaluated
many times to search the parameter space; see Appendix \ref{app:snmode})
takes approximately 45 minutes for the $N=2928$ SK dataset on a DEC
Alpha 250/ev5 and roughly a factor of five less on a Cray J90 parallel
supercomputer; compressing to 1200 eigenmodes takes only five minutes on
the DEC including overhead from the compression process.  Upcoming
balloon datasets are expected to have at least an order of magnitude
more data---which translates to a factor of 1000 in execution time (and
100 in storage requirements). Megapixel datasets foreseen for upcoming
satellite missions are clearly too large to analyze in this way with any
foreseeable increase in computer speed.

The quadratic estimator is also $O(N^3)$ despite claims that it is
$O(N^2)$\cite{tegmarkoptimal}.  Finding good approximations that will
reduce it to $O(N^2)$ is an unsolved problem, crucial for further
study.

Even as we have implemented it, the quadratic is much faster than direct
evaluation of the likelihood function.  Starting from the
signal-to-noise basis, one iteration of the quadratic estimator for the
10 SK bands of \S~\ref{sec:RadComp} took 250 seconds to calculate the
window function rotated into that basis, and 180 seconds to form the
Fisher matrix and calculate the quadratic estimator on the DEC Alpha,
compressing to 1200 modes. The direct evaluation method, in contrast,
requires a new rotation to the signal-to-noise basis at each band, which
is roughly 5 minutes per band, using the same 1200-mode compression.

We have also performed the quadratic calculation via direct
evaluation of $\delta a_p$ and the Fisher matrix in the pixel basis,
calculating quantities like $C^{-1}C_{T,p}$ using the Cholesky
decomposition of $C$. This is somewhat faster than the same calculation
in the eigenmode basis, although it does not allow easy implementation
of signal-to-noise compression.

In Appendix \ref{app:snmode} we explicitly calculate the Fisher matrix
in $O(N^3)$ operations (the signal-to-noise eigenmode decomposition). To
see what makes the quadratic estimator an $O(N^3)$ operation in general,
it helps to rewrite it.  If we define
\be
y_p \equiv \Delta^T C^{-1}C_{T,p} C^{-1} \Delta
\ee
then $\langle y_p \rangle = {\rm Tr}(C^{-1} C_{T,p})$ and
we can rewrite the quadratic estimator as
\be
\delta a_p = {1\over 2}\sum_{p'}\left(F^{(a)}\right)^{-1}_{pp'}\left(y_{p'} - 
\langle y_{p'} \rangle\right)
\ee
We can iteratively solve for the vector $C^{-1}\Delta$ and therefore
$y_p$ can be calculated.  The slowest parts of the quadratic are
the Fisher matrix and $\langle y_p \rangle$---both of which
require calculating $C^{-1}C_{T,p}$.

If we can find a good approximation to $C^{-1}C_{T,p}$ that can be
calculated in $O(N^2)$ operations, then the entire estimation procedure
will be $O(N^2)$ for each element of the Fisher matrix.  Since the
Fisher matrix has $N_p^2$ elements, the estimation procedure is $O(N^2
N_p^2)$.  If the number of parameters is roughly the square root of the
number of pixels (as is expected to be the case for power spectrum
estimation) then the estimation procedure is $O(N^3)$.  For the largest
maps, we can take advantage of the sparseness of the Fisher matrix to
only calculate it in a band around the diagonal, reducing the process to
$O(N^{2.5})$.  [If we skip the power spectrum and go straight to the
estimation of cosmological parameters, then $N_p \ll N$ and the process
is $O(N^2)$.  Of course, if $C^{-1}$ is calculated directly (an $O(N^3)$
operation), then this is the most intensive step in calculating the
Fisher matrix, and the whole process is still $O(N^3)$.]
                                
The method we outline is completely general, allowing arbitrary chopping
strategies and off-diagonal noise correlations, including those
generated by the subtraction of constraints or foreground templates, as
explained in \S~\ref{sec:methodsLike} and Appendix 
\ref{app:snmode}. We expect that these noise correlations will become
increasingly important in future balloon and satellite experiments,
which will exhibit both $1/f$ streaking and significant foreground
contamination. Although we hope to find techniques that will reduce the
computational load from $O(N^3)$ to $O(N^2)$, with general 
inhomogeneous noise this is a difficult problem. One approach is to try
to find the best possible approximation to the generalized noise matrix
which allows fast computation, then treat the residual perturbatively.
Another is to rely on the special nature of the noise for a given
experiment. For example, if an approximate set of eigenmodes along with their
projection onto the spherical harmonics is known for the geometry and
weighting of a particular dataset, then quantities like $C^{-1}C_{T,p}$ can be
calculated without explicit inversion or matrix manipulation. Gorski's
cut-sky spherical harmonics\cite{gorskiCl} have this property, but
require an $O(N^3)$ Cholesky decomposition for their construction.

For mapping experiments, the parameter derivatives $C_{T,\ell}$ will
be proportional to the Legendre polynomials, which can in turn be
written as a sum over spherical harmonics using the appropriate
summation formula. 
We have shown that at high $\ell$,
two-dimensional (flat-sky) fourier modes with wavenumber $\vert {\bf
Q}\vert \sim \ell$ are very useful, and expect that they will be
effective as we look for ways to improve the computational speed.  For
COBE/DMR, using an approximate weight is adequate for some statistical
measures, but for high precision work the residual $60^\circ$
correlation and constraints should be taken into account. For upcoming
balloon and satellite experiments, full and correct modelling of the
noise and its behaviour in various subspaces will be essential for
achieving the forecasted accuracy\cite{forecast,bet97} in cosmological
parameter determinations. 

\subsection{Redshift Surveys}

So far, we have concentrated our analysis on applications to CMB
anisotropy data. However, much of it can be carried over to estimate the
power spectrum of other sorts of data, particularly that of upcoming
redshift surveys\cite{galaxysurveys,THSVS}. In that case, we partition the
three-dimensional volume probed by the surveys into bins $i=1\ldots N$
and use counts-in-cells as the data $\Delta_i=s_i+n_i$.  Now, the ``beam
function'' becomes the selection function of the survey restricted to
the individual bins, which accounts for the flux cutoff in its
observational bands. The noise becomes considerably more complicated: it
is the ``shot noise'' which comes from the sampling of the underlying
density field in whose correlations we are actually interested.  This
noise is not Gaussian, but Poissonian (and only that if we ignore
correlations within the bin); to use this formalism requires that we
have enough galaxies per bin that a Gaussian approximation is adequate,
but small enough that the correlations within the bin are ignorable (and
small enough that we still have information on scales of interest). In
that case, the Poisson noise term has $\langle n_i^2 \rangle$ given by
the counts in the bin.  Of course, there are further complications due
to redshift-space distortions. For an alternative to this procedure, see
\cite{DodHuiJaf}.

\subsection{Summary}

We have demonstrated two techniques for determining the power spectrum
of CMB fluctuations from realistic microwave data. We have presented an
analysis of both a direct likelihood search and a specific quadratic
estimator; the most important result of this paper is the proof that the
iterated application of the quadratic estimator is a fast method for
finding the peak and curvature of the likelihood function.

Our methods easily incorporate such realistic features as
convoluted chopping strategies, incomplete sky coverage, and the removal
of linear constraints from the data. As implemented today, our method
requires $O(N^3)$ operations in order to deal with these complications.
We have applied the techniques to both the DMR and SK datasets, which
exhibit all of these complications. Numerically, our results agree quite
well with other analyses of these datasets.

We have also discussed several caveats in the further use of the power
spectrum, associated with the non-Gaussian nature of the posterior
distribition of the ${\cal C}_\ell$. This can have repercussions in any
analysis (such as $\chi^2$, or even in our own rebinning techniques)
which implicitly or explicitly assume Gaussianity of the distribution
({\it i.e.}, the constant curvature of the log-likelihood).

The traditional procedure for reporting constraints on the power
spectrum is the band-power method, where the power spectrum estimate
is considered to be a measurement of the power averaged through some
specific filter.  In the past this filter has been given by the trace
of the window function, ${W}_\ell$.  We advocate a generalization of this
procedure where the filter is derived from the Fisher matrix instead.
With this better definition of the filter, the new technique will
improve the accuracy of analyses that start from band-power estimates.

\subsection{Quadratic Estimation Cookbook}

We now summarize the complete algorithm for quadratic power
spectrum estimation: 
\begin{enumerate}
\item Obtain the data and the error or weight matrix $C_N$ (including
  the effects of constraints as discussed in Appendix \ref{app:snmode}).
\item Choose an initial $\ell$ binning, as discussed in
  \S~\ref{sec:binning}.
\item Calculate the window function matrix $W_{pp'}(\ell)$,
  Eq.~\ref{eqn:wl}, perhaps averaged over the $\ell$ bins.
\item Choose a power spectrum ${\cal C}_\ell^{(0)}$ to begin the
  iteration.
\item Calculate $C_T$ for ${\cal C}^{(i)}_\ell$ ($i=0$ for the first
  iteration), from Eq.~\ref{eqn:CT}
\item If desired, the rest of the calculation can be performed in the
  Signal-to-Noise basis of Appendix \ref{app:snmode}. In that case,
  $C_T$ and the data are transformed according to
  Eqs.~\ref{eqn:SNtrans1}--\ref{eqn:SNtrans2}.
\item Calculate the parameter derivatives $C_{T,B}\equiv\partial
  C_T/\partial q_B$ in each band, using
  Eqs.~\ref{eqn:CTell}--\ref{eqn:CTB} or, in the S/N basis,
  Eqs.~\ref{eqn:wkkl}--\ref{eqn:wkkB}. The parameter $q_B$,
  Eq.~\ref{eqn:Clparama}, is the fractional difference from
  $\C^{(i)}_\ell$.
\item Calculate the Fisher Matrix, Eq.~\ref{eqn:fish} or
  Eq.~\ref{eqn:SNfish}, for the chosen bands.
\item Calculate the complete quadratic $\delta q_B$ using
  Eq.~\ref{eqn:quadest} or Eq.~\ref{eqn:SNquadest}, and set \[{\cal
    C}_\ell^{(i+1)}=\sum_B (1+\delta q_B){\cal
    C}_\ell^{(i)}\chi_B(\ell).\]
\item Lather, rinse, and repeat with Step 5 until $\delta q_B\approx 0$
  to the desired accuracy.
\end{enumerate}

This description has not included the complications associated with
rebinning (see \S~\ref{sec:binning}) and the use of filter functions
for reporting bandpowers (see \S~\ref{sec:RadComp}).

\subsection{Numerical Results}
Our power spectrum estimates for COBE/DMR and SK are available over the WWW
and by anonymous FTP in the directory
{\tt file://ftp.cita.utoronto.ca/\linebreak[0]cita/knox/pspec\_Cl/}. These numerical
results include the results of both the full-likelihood and quadratic
procedures; for the latter we include the results for 
``orthogonalized'' and ``shaped'' bands, along with appropriately
tabulated filter functions.

\acknowledgements The authors thank Scott Dodelson, Andrew Hamilton,
Uros Seljak and Max Tegmark for useful conversations, and Ted Bunn and
Kris Gorski for providing the results of their analyses. AJ and LK would
especially like to acknowledge the hospitality of the Aspen Center for
Physics, where portions of this work were completed. Some of the
computations described herein were performed by AJ on the computers of
the National Energy Research Scientific Computing Center (NERSC).

\appendix
\section{Signal-to-Noise Eigenmodes \& Constraints}
\label{app:snmode}

Some of the calculations described in this paper are performed in the
``signal-to-noise eigenmode'' basis\cite{BondJaffe,Bond,Bond94,BunnWhite}. To
effect this transformation, we model the observation at a pixel as
\begin{equation}
        \Delta_i=s_i + n_i
\end{equation}
where $s_i$ is the contribution to the signal, and $n_i$ to the noise.
They have zero means, and independent correlation matrixes
$\VEV{n_in_{i'}}=C_{nii'}$ and $\VEV{s_is_{i'}}=\sigma_{\rm th}^2 C_{Tii'}$.
Here, $\sigma_{\rm th}$ is the unknown amplitude of the signal to be
measured (along with other possible parameters in $C_T$).

We may ascribe more than the experimental noise contribution to $n_i$:
in particular, any contributions to the observation with which we are
not concerned in a given part of the calculation can be included in the
noise. This could be the CMB monopole and dipole, or constraints such as
averages and gradients that may have been removed from the data to
compensate for atmospheric and instrumental drift.
For COBE/DMR, we allow arbitrary amplitudes for the monopole (one
component) and the dipole (three components); for SK, we allow an
arbitrary average for each ``demodulation''\cite{nett95}, giving a total
of 66 separate amplitudes.  In the event, each constraint component $c$
can be represented by a template in pixel space, $\Upsilon_{ci}$, with
an unknown amplitude, $\kappa_c$. Thus, the CMB signal plus experimental
noise is given by the combination $\Delta_i - \sum_c\kappa_c
\Upsilon_{ci}$, which is distributed as a Gaussian with correlation
matrix $C_n+{\sigma_{\rm th}^2}C_T$. We do not know the amplitudes
$\kappa_c$ {\em a priori}, but we can assign them a prior probability
distribution given by a zero-mean Gaussian with very large variances in
the matrix $\langle\kappa_c\kappa_{c'}\rangle=K_{cc'}$, (compared to the
expected signal and the experimental noise), and then marginalize over
the amplitudes $\kappa_c$. It turns out that this marginalization
procedure can be done analytically, and the result is that the
likelihood is now given by a zero-mean Gaussian distribution in $\Delta$
alone, with a full correlation matrix including a new term accounting
for the unknown constraints:
\begin{equation}
  \VEV{\Delta_i \Delta_{i'}} =
  \sigma_{\rm th}^2 C_{Tii'} + C_{nii'} + C_{Cii'}
\end{equation}
where
\begin{equation}
C_{Cii'} = \sum_{cc'}\Upsilon_{ci} K_{cc'}\Upsilon_{c'i'}
\end{equation}
is the constraint or template correlation matrix. For a diagonal matrix
of priors, $K={\rm diag}(\sigma_c^2)$, this reduces to $C_{Cii'}=\sum_c
\sigma_c^2 \Upsilon_{ci}\Upsilon_{ci'}$.

In effect, we have added a new term to the noise correlation,
$C_N=C_n+C_C$; in the following we shall implicitly include this in
$C_N$.  In the limit $\sigma_c^2\to\infty$, this procedure is equivalent
to projecting out the constrained components from the data and the
correlation matrix; because this projection results in a singular matrix, the
marginalization procedure is numerically simpler (but see \cite{THSVS}
for the details of an implementation of the projection procedure).  Note also that this
procedure is more generally useful: in particular it provides a new
technique for removing foreground contamination with a known spatial
morphology\cite{JaffeFG}.

With this split of the observation into signal and (generalized) noise,
we first perform a so-called whitening transformation
\begin{eqnarray}\label{eqn:SNtrans1}
        C_N &\to& C_N^{-1/2} C_N  C_N^{-1/2} = I;\nonumber\\
        C_T &\to& C_N^{-1/2} C_T  C_N^{-1/2};\nonumber\\
        \Delta&\to& C_N^{-1/2} \Delta.
\end{eqnarray}
Here, $C_N^{-1/2}$ is the inverse of the Cholesky decomposition of $C_N$ 
or its Hermitian Square Root.
Now, the noise part of the ``new data,'' $C_N^{-1/2}\Delta$, are uncorrelated, 
with unit variance. We next diagonalize the signal matrix with its 
matrix of eigenvectors, 
\begin{eqnarray}\label{eqn:SNtrans2}
        C_T&\to& R^\dagger C_N^{-1/2} C_T  C_N^{-1/2} R=
        {\cal E}={\rm diag}({\cal E}_k);       \nonumber\\
        \Delta&\to& R^\dagger C_N^{-1/2} \Delta = \xi.
\end{eqnarray}
In this new basis, the data $\xi_k$ are uncorrelated with variance
$\VEV{\xi_k^2}=1+\sigma_{\rm th}^2{\cal E}_k$. The ${\cal E}_k$ are
``eigenmodes of signal-to-noise''; modes with large eigenvalue are
expected to be well-measured (for the specific theory matrix $C_T$ used
in the transformation); modes with small eigenvalue are poorly-measured
(and do not contribute significantly to the likelihood).  In particular,
we use this transformation to compress the SK data: we pick a fiducial
model (in this case, $n_s=1.45$ tilted standard CDM, which fits the SK
data alone reasonably well) and calculate the modes for this theory. We
then discard all but the top 1200 modes (of 2928 data points) and treat
this linear combination as our new dataset (for which we subsequently
calculate all likelihoods without further approximation);
elsewhere\cite{BondJaffe} we show that this truncation to 1200
theory-dependent modes is an excellent approximation to the entire 
dataset.

Note that in the S/N basis, the likelihood as a function of the 
amplitude $\sigma_{\rm th}$ is quite easy to compute for arbitrary 
values:
\begin{equation}
  -2\ln P(\Delta|\sigma_{\rm th}^2C_\ell) =  \sum_k \left[
    \ln(1+\sigma_{\rm th}^2{\cal E}_k) + 
    {\xi_k^2\over1+\sigma_{\rm th}^2{\cal E}_k}\right]
\end{equation}
(up to a constant).
In the calculation of the likelihood as a function of the values of 
the power spectrum, we iterate by ascribing only the single $C_\ell$ 
(or within a band, with some shape for $C_\ell$ over the band) 
of interest to the signal, $s_i$, and the rest to the noise, $n_i$, 
along with the actual experimental noise, and any terms due to 
constraints such as dipole removal. This way, the single parameter of 
interest at anytime is just the amplitude $\sigma_{\rm th}^2\propto 
{\cal C}_\ell$ for that band, for which the likelihood is easy to compute 
once the S/N mode decomposition has been determined.

We also compute the quadratic $C_\ell$ estimators in this basis. First,
we define the window function matrix (Eq.~\ref{eqn:wl}) transformed
into this basis,
\begin{equation}\label{eqn:wkkl}
  G_{kk'}(\ell) = \sum_{ii'}
  \left(R^\dagger C_N^{-1/2}\right)_{ki} W_{ii'}(\ell) 
  \left(C_N^{-1/2} R\right)_{i'k'}.
\end{equation}
This quantity comes into the calculations because it is related to the
derivative of the theory covariance in the eigenbasis,
\begin{eqnarray}\label{eqn:wkkB}
  {\cal E}_{kk',\ell}&\equiv& \sum_{ii'}
  \left(R^\dagger C_N^{-1/2}\right)_{ki}
  {\partial C_{Tii'}\over\partial{\cal C}_\ell} \left(C_N^{-1/2}
    R\right)_{i'k'}\nonumber\\
  &=&{{\ell+1/2}\over{\ell(\ell+1)}}G_{kk'}(\ell).
\end{eqnarray}
Here we have assumed that we are interested in the individual ${\cal
  C}_\ell$ values. If we are instead interested in the values over some
bands, $B$, of $\ell$ with some assumed spectral shape ${\cal
  C}_{\ell}^{\rm shape}$, then we use instead 
\begin{equation}
  {\cal E}_{kk',B}=\sum_{\ell\in B} 
  {\cal E}_{kk',\ell}{\cal C}_{\ell}^{\rm shape}.
\end{equation}
Note that, unlike the full theory covariance, ${\cal E}={\rm diag}({\cal E}_k)$, these
derivatives have off-diagonal components. In
Eqs.~\ref{eqn:SNfish}--\ref{eqn:SNquadest} below, ${\cal E}_{kk',B}$ and
${\cal E}_{kk',\ell}$ can be used interchangeably, depending on whether
one is estimating individual ${\cal C}_\ell$ values, or those in bands.

The Fisher matrix for the parameters ${\cal
  C}_\ell$ (Eq.~\ref{eqn:fish}) then becomes
\begin{equation}\label{eqn:SNfish}
  F_{\ell\ell'}=
  \sum_{kk'}{{\cal E}_{kk',\ell}{\cal E}_{k'k,\ell'}\over
    (1+\sigma_{\rm th}^2{\cal E}_k)(1+\sigma_{\rm th}^2{\cal E}_{k'})}.
\end{equation}
Then the full quadratic estimator (compare Eq.~\ref{eqn:quadest})
is
\begin{eqnarray}\label{eqn:SNquadest}
  \delta {\cal C}_\ell = &&  {1\over2}\sum_{\ell'}F_{\ell\ell'}^{-1}
 \Bigg[\nonumber\\
  &&\sum_{kk'}
  {\xi_k {\cal E}_{kk',\ell'}\xi_{k'}\over   
    (1+\sigma_{\rm th}^2{\cal E}_k)(1+\sigma_{\rm th}^2{\cal E}_{k'})}
  -\sum_k
  {{\cal E}_{kk,\ell'}\over1+\sigma_{\rm th}^2{\cal E}_k}\Bigg].\nonumber\\&&
\end{eqnarray}
Note that in this formalism the $O(N^3)$ transformation into the S/N
basis is the most expensive part of the calculation; the remainder
requires trivial $O(N^2)$ sums and the inverse of the (comparitively small)
$N_p\times N_p$ Fisher matrix. 

\section{Rebinning Orthogonal Linear Combinations}
Here we derive Eq.~\ref{eqn:calCbeta} which tells how to
rebin orthogonal linear combinations of ${\cal C}_\ell$.
We then generalize to the case where the initial binning is
coarser than $\Delta \ell = 1$.

We start by parameterizing the spectrum as 
\be
{\cal C}_\ell = q_\ell {\cal C}_\ell^{\rm shape}
\ee
and then transforming the $q_\ell$ to 
$\tilde q = Z^T q$.  If we assume the shape is
correct, then the expectation value of
$\tilde q_\lambda^N \equiv \tilde q_\lambda/N_\lambda$
is independent of $\lambda$, where $N_\lambda = \sum_\ell 
Z_{\ell\lambda}$.  Since we always want to average  things together that
we expect to be measurements of the same quantity, we
average together the $\tilde q_\lambda^N$.  Calling the result
$\tilde q_\beta^N$:
\be
\tilde q_\beta^N = {\sum_{\lambda \lambda'} \tilde q_\lambda^N
F_{\lambda \lambda'}^{\tilde q^N} \over \sum_{\lambda \lambda'} 
F_{\lambda \lambda'}^{\tilde q^N}}.
\ee
Here, and in the following, the sums over $\lambda$ and $\lambda'$
extend only over the range determined by $\beta$.  For example,
if for $\beta = 1$ we are averaging together the first three linear 
combinations, then the sums over $\lambda$ and $\lambda'$ run 
from one to three.

Using the fact that $F_{\lambda \lambda'}^{\tilde q^N} = N_\lambda
F_{\lambda \lambda'}^{\tilde q} N_{\lambda'}$ and specializing 
to the case where $F_{\lambda \lambda'}^{\tilde q} = 
\delta_{\lambda \lambda'}$ (which is the case for $Z = L$ or
$Z = F^{1/2}$) we get
\be
\tilde q_\beta^N = {\sum_\lambda \tilde q_\lambda 
N_\lambda \over \sum_\lambda N_\lambda^2}.
\ee
Plugging in $\tilde q_\lambda = Z_{\ell\lambda} 
{\cal C}_\ell/{\cal C}_\ell^{\rm shape}$ and $N_\lambda = \sum_\ell
Z_{\ell\lambda}$ a little algebra shows that the filter function
is 
\be
f^{(\C)}_{\beta \ell} = \sum_\lambda f^{(\C)}_{\lambda \ell} N_\lambda
\ee
where
\be
f^{(\C)}_{\lambda \ell} = Z_{\ell\lambda}/{\cal C}_\ell^{\rm shape}
\ee
is the filter function prior to rebinning.  Therefore
\be
{\cal C}_\beta = {\sum_\lambda \tilde q_\lambda N_\lambda
\over \sum_\ell f^{(\C)}_{\beta \ell} } = {\sum_\ell q_\ell \C_\ell^{\rm shape}
f^{(\C)}_{\beta \ell} \over \sum_\ell f^{(\C)}_{\beta \ell}}
\ee
which is Eq.~\ref{eqn:calCbeta}.  

As an aside, we consider the case of rebinning all the estimates
into one bin.  We expect that the estimated power and filter
function in this case should be independent of the basis of the
original estimates; they should not depend on $Z$.  Indeed, this
is the case:
\be
\tilde q_\beta^N = {\sum_{\ell \ell' \lambda} {\cal C}_\ell/{\cal C}_\ell^{\rm shape}
Z_{\ell\lambda}Z_{\ell' \lambda} \over
\sum_{\ell \ell' \lambda} Z_{\ell\lambda} Z_{\ell' \lambda}}
= {\sum_{\ell \ell'} {\cal C}_\ell/{\cal C}_\ell^{\rm shape} F^q_{\ell \ell'} \over
\sum_{\ell \ell'}  F^q_{\ell \ell'} }.
\ee
The second equality follows since when the sum over $\lambda$
goes over all $\lambda$, $Z_{\ell\lambda} Z^T_{\lambda \ell'} = F^q_{\ell\ell'}$.

When the initial binning is coarser than $\Delta \ell = 1$, then this
procedure is slightly more complicated.  We introduce the sub-band
structure filter, $f^{(\C)}_{B\ell}$, which is defined within each band $B$.
The sub-band structure is given by 
\be
f^{(\C)}_{B\ell} = \sum_{\ell'} F_{\ell \ell'}^{(\C)}\C_{\ell'}^{\rm shape}
\ee
which is the same as Eq.~\ref{eqn:fBl}.  The difference here
is that we have not calculated $F_{\ell \ell'}^{(\C)}$ and thus must
rely on analytic knowledge of it.  

The rebinning procedure is the same for the $\Delta \ell = 1$ initial
binning except
\bea
\C_\ell^{\rm shape} \rightarrow \C_B^{\rm shape} &=& 
{\sum_{\ell \in B} f^{(\C)}_{B\ell} \C_\ell^{\rm shape} \over
\sum_{\ell \in B} f^{(\C)}_{B\ell}},\\
N_\lambda \rightarrow N_\beta &=& \sum_B Z_{B\beta} 
\eea
and
\be
f^{(\C)}_{\beta \ell} \rightarrow f^{(\C)}_{\beta' \ell} = \left({f^{(\C)}_{B\ell}  \over
\sum_{\ell \in B} f^{(\C)}_{B\ell}}\right)f^{(\C)}_{\beta' B(\ell)}
\ee
where
\be
f^{(\C)}_{\beta' B} = \sum_{\beta \in \beta'} {Z_{B\beta} \over \C_B^{\rm shape}}
N_\beta.
\ee
The final result is 
\be
\C_{\beta'} = {\sum_\ell q_{B(\ell)} \C_\ell^{\rm shape}f^{(\C)}_{\beta'\ell} \over
\sum_\ell f^{(\C)}_{\beta' \ell}} = \sum_B X_{\beta' B} q_B.
\ee
The last equality is used to define the matrix $X_{\beta' B}$ and to
emphasize that $\C_{\beta'}$ is simply a linear transformation of the original
$q_B$ parameters.  The
Fisher matrix for $\C_{\beta'}$ can easily be calculated from that
for $q_B$, using the general
rule for how the Fisher matrix changes under linear transformation of the
parameters.


\begin{thebibliography}{ucsc}
\bibitem{knox95} L.\ Knox, {\sl Phys.\ Rev.}\/ {\bf D52} 4307 (1995).
\bibitem{forecast} G.\ Jungman, M.\ Kamionkowski,
  A.\ Kosowsky \& D.N.\ Spergel, {\sl Phys.\ Rev.\ Lett.}\/ {\bf 76},
  1007 (1996); {\sl ibid},  {\sl Phys.\ Rev.\ Lett.}\/ {\bf D54},
  1332 (1996); M.\ Zaldarriaga, D.\ Spergel \& U.\ Seljak,
  astro-ph/9702157.
\bibitem{bet97} J.R.\ Bond, G.\ Efstathiou \& M.\ Tegmark, 
  astro-ph/9702100.
\bibitem{ourotherpaper} A.H.\ Jaffe, L.\ Knox \& J.R.\ Bond, in preparation.
\bibitem{jkb} A.H.\ Jaffe, L.\ Knox \& J.R.\ Bond, {\sl Proceedings of
    the Eighteenth Texas Symposium on Relativistic Astrophysics}, ed.\ 
   A.\ Olinto, J.\ Frieman \& D.\ Schramm, Chicago, IL, 1996,
  World Scientific, in press.
\bibitem{HauPeebWright} M.G.\ Hauser \& P.J.E.\ Peebles, 
{\sl Ap.\ J.}\/ {\bf 185}, 757 (1973); E.L.\ Wright, 
{\sl Ap.\ J.}\/ {\bf 436}, 443 (1994).
\bibitem{tegmarkoptimal} M.\ Tegmark, astro-ph/9611174, 
  {\sl Phys.\ Rev.}\/ {\bf D55}, 5895 (1997).
\bibitem{kbj} L.\ Knox, J.R.\ Bond \& A.\ Jaffe, in \cite{jkb}
\bibitem{DMR} C.L.\ Bennett, A.J.\ Banday,   K.M.\ Gorski, G.\ Hinshaw, 
 P.D.\ Jackson, P.\ Keegstra, A.\ Kogut, G.F.\ Smoot,D.\
  Wilkinson \& E.L.\ Wright,  {\sl Ap.\ J.\ Lett.}\/ {\bf 464}, L1
  (1996), and 4-year DMR references therein. 
\bibitem{dmrFGs} A.\ Kogut, A.J.\ Banday, C.L.\ Bennett, K.M.\ Gorski,
  G.\ Hinshaw, G.F.\ Smoot \& E.L.\ Wright, {\sl Ap.\ J.\ Lett.}\/ {\bf
    464}, L5 (1996); A.\ Kogut, A.J.\ Banday, C.L.\ Bennett, K.M.\ 
  Gorski, G.\ Hinshaw, P.D.\ Jackson, P.\ Keegstra, C.\ Lineweaver,
  G.F.\ Smoot, L.\ Tenorio \& E.L.\ Wright, {\sl Ap.\ J.}\/ {\bf 470},
  653 (1996); A.J.\ Banday, K.M.\ Gorski, C.L.\ Bennett, G.\ Hinshaw,
  A.\ Kogut \& G.F.\ Smoot, {\sl Ap.\ J.\ Lett.}\/ {\bf 468} L85 (1996).
\bibitem{nett95} C.B.\ Netterfield, M.J.\ Devlin, N.\ Jarosik, L.\ Page \&
  E.J.\ Wollack,  {\sl Ap.\ J.}\/ {\bf 474}, 47 (1997); E.J.\ Wollack,
  M.J.\ Devlin, N.\ Jarosik, C.B.\ Netterfield, L.\ Page \& D.\
  Wilkinson, {\sl Ap.\ J.}\/ {\bf 476}, 440 (1997).
\bibitem{Lineweaver} C.\ Lineweaver \& D.\ Barbosa, astro-ph/9706077.
\bibitem{BondJaffe} J.R.\ Bond \& A.\ Jaffe, {\sl Phys.\ Rev.}\/ {\bf
    D}, submitted; {\sl ibid}, in {\sl Proceedings of the
  XVIth Moriond meeting, ``Microwave Background Anisotropies,''} ed.\
  F.R.\ Bouchet \etal\ (Gif-Sur-Yvette: Editions Fronti\`eres) (1997).
\bibitem{HancockRocha} S.\ Hancock and G.\ Rocha, astro-ph/9612016; 
{\sl ibid}, in F.R.\ Bouchet \etal, {\sl op~cit}.
\bibitem{Bond} J.R.\ Bond, {\sl Astrophys.\ Lett.\ Comm.}\/ {\bf 32}, 63
  (1995).
\bibitem{Bond94} J.R.\ Bond, {\sl Phys.\ Rev.\ Lett.}\/ {\bf 74}, 4369.
\bibitem{BunnWhite} E.F.\ Bunn \& M.\ White,  {\sl Ap.\ J.}\/ {\bf 480},
  6 (1997).
\bibitem{TTH} For a review of signal-to-noise eigenmodes see
M.\ Tegmark, A.M.\ Taylor \& A.F.\ Heavens, {\sl Ap.\ J.}\/ {\bf
480}, 22 (1997).
\bibitem{THSVS} M.\ Tegmark, A.J.S.\ Hamilton, M.A.\ Strauss, M.S.\
Vogeley \& A.S.\ Szalay, astro-ph/9708020.
\bibitem{tegmarkdeltal} M.\ Tegmark, {\sl
Mon. Not. R. Astronom. Soc.}, 280, 299 (1995); 
astro-ph/9705188, {\sl Phys.\ Rev.}\/ {\bf D}, in press.
\bibitem{numrec} W.H.\ Press, B.\ Flannery, S.A.\ Teukolsky \& W.T.\
  Vetterling, 
{\sl Numerical Recipes in Fortran, 2nd ed.}, 
(Cambridge:  Cambridge Univerisity Press) (1992). 
\bibitem{Hamilton} A.J.S.\ Hamilton, astro-ph/9701008; astro-ph/9701009.
\bibitem{TegHam}  M.\ Tegmark \& A.\ Hamilton, in J.\ Frieman \etal,
  {\sl op~cit}.
\bibitem{BCstat}
S.P.\ Boughn, E.S.\ Cheng, D.A.\ Cottingham, D.J.\  Fixsen, {\sl  Ap.\
J.\ Lett.}\/ {\bf 391}, L49 (1992). 
\bibitem{gorskiCl} K.M.\ Gorski, {\sl Ap.\ J.\ Lett.}\/ {\bf 430} L85
  (1994); K.M.\ Gorski, A.J.\
  Banday, C.L.\ Bennett, G.\ Hinshaw,  A.\ Kogut, G.F.\ Smoot \& E.L.\
  Wright, {\sl Ap.\ J.\ Lett.}\/ {\bf 464} L11 (1996); K.M.\ Gorski,
  astro-ph/9701191.
\bibitem{LineSmoot} C.\ Lineweaver \& G.\ Smoot, COBE not 5051 (1993).
\bibitem{GneSmoot} R.\ Kneissl \& G.\ Smoot, COBE note 5053 (1993).
\bibitem{Leitch} E.\ Leitch, private communication (1997).
\bibitem{nettPriv} C.B.\ Netterfield \& L.\ Page, private communication.
\bibitem{bh95} Bond, J.R., in {\it Cosmology and Large Scale
Structure}, Les Houches Session LX, August 1993, ed. R. Schaeffer
\etal\ (Elsevier Science Press, Amsterdam), pp. 469-674 (1996).
\bibitem{SKmap} A map of the North Celestial Pole region has
been made by Wiener-filtering the SK data:  M.\ Tegmark, A.\ de 
Oliveira-Costa, M.J.\ Devlin,  C.B.\ Netterfield, L.\ Page \& 
E.J.\ Wollack, {\sl Ap.\ J.\ Lett.}\/ {\bf 474} L77 (1997). See also \cite{JaffeFG}
\bibitem{map96ref} C.\ Bennett \etal, MAP experiment home page,\\ 
{\tt http://map.gsfc.nasa.gov} (1996).
\bibitem{cobrassamba96ref} M.\ Bersanelli \etal, {\sl COBRAS/SAMBA,
The Phase A Study for an ESA M3 Mission}, preprint (1996); Planck home
page, {\tt
  http://astro.estec.esa.nl/SA-general/\linebreak[0]Projects/Cobras/cobras.html} (1996). 
\bibitem{TopHat97}
{E.\ Cheng \etal, TopHat home page,\\ 
{\tt http://\linebreak[1]cobi.gsfc.nasa.gov/msam-tophat.html} (1997).}
\bibitem{MAXIMA97}{P.\ Richards \etal, MAXIMA home page,\\ 
{\tt http://physics7.berkeley.edu/group/cmb/gen.html}
(1997).} 
\bibitem{Boomerang97}
{A.\ Lange \etal,
Boomerang home page,\\ {\tt http://\linebreak[1]astro.caltech.edu/\~{}mc/boom/boom.html} (1997).}
\bibitem{ACE97} P.\ Lubin \etal, ACE/Beast home page,\\ 
{\tt http://www.deepspace.ucsb.edu/research/\linebreak[0]Sphome.html} (1997). 
\bibitem{galaxysurveys} J.E.\ Gunn \& D.H.\ Weinberg, in {\it Wide-Field
Spectroscopy and the Distant Universe}, ed.\ S.J.\ Maddox \& A.\
Arag\`on-Salamanca (Singapore: World Scientific), p.\ 3 (1995); M.\
Tegmark, astro-ph/9706198. 
\bibitem{DodHuiJaf} S.\ Dodelson, L.\ Hui \& A.H.\ Jaffe, in preparation.
\bibitem{JaffeFG} A.H.\ Jaffe, D.\ Finkbeiner \& J.R.\ Bond, in preparation.
\end{thebibliography}
\end{document}